\documentclass[review]{elsarticle}
\usepackage{amssymb,amsthm,mathrsfs,graphicx, bm}
\usepackage{natbib}
\usepackage{subcaption}
\usepackage{titlesec}
\usepackage{geometry}
\usepackage{fancyhdr}
\usepackage{newtxtext,newtxmath}
\usepackage{hyperref}

\usepackage{amsmath}
\usepackage{color}
\usepackage{amsfonts}
\usepackage{pgfplots}
\pgfplotsset{compat=newest}
\usetikzlibrary{plotmarks}
\usetikzlibrary{arrows.meta}
\usepgfplotslibrary{patchplots}
\usepackage{grffile}

\numberwithin{equation}{section}

\titlespacing{\section}{0pt}{\parskip}{0.01\parskip}

\usepackage{lineno,hyperref}
\usepackage{moreverb}
\usepackage{bm,color,soul,transparent,amsmath, amsmath,amssymb,amsthm,fancyhdr,mathrsfs,graphicx,setspace,stmaryrd,psfrag}
\usepackage[ruled,slide]{algorithm2e}
\newcommand\BibTeX{{\rmfamily B\kern-.05em \textsc{i\kern-.025em b}\kern-.08em
T\kern-.1667em\lower.7ex\hbox{E}\kern-.125emX}}

\journal{CMAME}

\begin {document}

\begin{frontmatter}

\title{Numerical investigation into fracture resistance of bone following adaptation}

\author[gla,vet]{Karol Lewandowski\corref{cor}}
\ead{karol.lewandowski@glasgow.ac.uk}
\author[gla]{{\L}ukasz Kaczmarczyk}
\ead{lukasz.kaczmarczyk@glasgow.ac.uk}
\author[gla]{Ignatios Athanasiadis}
\ead{ignatios.athanasiadis@glasgow.ac.uk}
\author[vet]{John F. Marshall}
\ead{john.f.marshall@glasgow.ac.uk}
\author[gla]{Chris J. Pearce}
\ead{chris.pearce@glasgow.ac.uk}
\cortext[cor]{Corresponding author}
\address[gla]{Glasgow Computational Engineering Centre, The James Watt School of Engineering, University of Glasgow, Glasgow, G12 8QQ, UK.}
\address[vet]{Weipers Centre Equine Hospital, School of Veterinary Medicine, University of Glasgow, Glasgow, G61 1QH, UK,}

\begin{abstract}
Bone adapts in response to its mechanical environment. 
This evolution of bone density is one of the most important mechanisms for developing fracture resistance. 
A finite element framework for simulating bone adaptation, commonly called bone remodelling, is presented. This is followed by a novel method to both quantify fracture resistance and to simulate fracture propagation. The authors' previous work on the application of configurational mechanics for modelling fracture is extended to include the influence of heterogeneous bone density distribution. The main advantage of this approach is that configurational forces, and fracture energy release rate, are expressed exclusively in terms of nodal quantities. This approach avoids the need for post-processing and enables a fully implicit formulation for modelling the evolving crack front.
In this paper density fields are generated from both (a) bone adaptation analysis and (b) subject-specific geometry and material properties obtained from CT scans. It is shown that, in order to correctly evaluate the configurational forces at the crack front, it is necessary to have a spatially smooth density field with higher regularity than if the field is directly approximated on the finite element mesh. Therefore, discrete density data is approximated as a smooth density field using a Moving Weighted Least Squares method.
Performance of the framework is demonstrated using numerical simulations for bone adaptation and subsequent crack propagation, including consideration of an equine 3\textsuperscript{rd} metacarpal bone. The degree of bone adaption is shown to influence both fracture resistance and the resulting crack path. 

\end{abstract}

\end{frontmatter}


\section{Introduction}
This paper presents a framework for the computational modelling of bone adaptation (commonly referred to as bone remodelling) and bone fracture, and their inter-relationship.
Bone adaptation is the on-going biological process of replacing old bone tissue with new bone, thus repairing fatigue damage~\citep{hughes2017role}.
This ability to repair bone micro-damage caused by cyclic loading is essential for maintaining mechanical integrity. Consequently, there is a strong correlation between stress fractures and the adaptation process~\citep{hughes2017role}. Furthermore, bone repair can be overwhelmed by load-induced bone densification that also increases brittleness and reduces fracture resistance~\citep{loughridge2017qualitative}.

One of the first mathematical theories for bone adaptation~\citep{cowin1976bone}, based on open system thermodynamics, has its foundation in the theory of poroelasticity. 
In this approach (unlike classical closed systems), energy, mass, momentum and entropy can be exchanged with the environment. It has been adopted and enhanced over the years
~\citep{harrigan1996bone, jacobs1995numerical, weinans1992behavior}.
This process of density evolution requires a~mechanical stimuli as a trigger for bone adaptation. This stimulus may take the form of stress~\citep{beaupre1990approach, carter1996mechanical, doblare2002anisotropic}, strains~\citep{cowin1976bone} or strain energy density~\citep{weinans1992behavior, kuhl2003theory,kaczmarczyk2011efficient, Connor2017bone}.

The use of computational tools to describe bone behaviour has gained a tremendous importance over the last decade. 
In particular, the Finite Element Method (FEM) has been used to improve understanding of the fracture behaviour of bones and the relationships between load conditions and bone architecture~\citep{podshivalov2014road, poelert2013patient}. However, there are only a few examples of the numerical analysis of both bone adaptation and fracture, e.g.~\citep{hambli2013integrated}. This paper presents a new computational framework based on FEM in order to predict bone density profiles 
(bone adaptation) due to exercise, quantify fracture resistance and simulate fracture propagation. This will improve understanding of the interrelationship between these phenomena and enable subject-specific simulations to be undertaken.

\begin{figure}[h]
	\centering
	   \begin{tikzpicture}
		\node at (0,4) {\includegraphics[width=14cm]{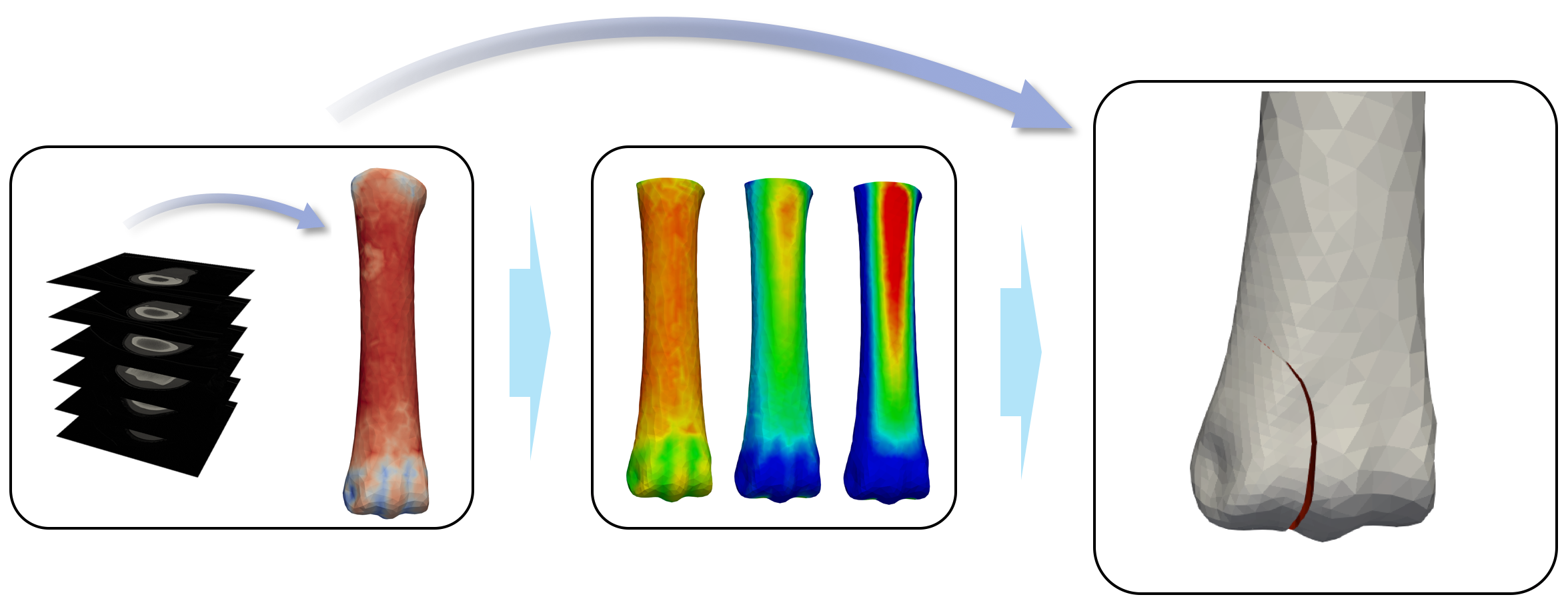}};
		\node at(0,0.6){\hspace{1cm} a) Density mapping. \hspace{1.3cm} b) Bone adaptation.  \hspace{1cm} c) Crack propagation analysis.  };
\end{tikzpicture}
\caption{Framework for estimating bone fracture resistance in MoFEM~\citep{mofem2017}. a) Density derived from Quantitative Computed Tomography (qCT) is mapped onto finite element mesh; (b) bone adaptation analysis; (c) assessment of fracture resistance and crack propagation analysis.}
\label{fig:framework}
\end{figure}

A schematic of the modelling framework is presented in Figure~\ref{fig:framework}. This paper extends the authors' previous work on modelling fracture propagation \citep{kaczmarczyk2014three,kaczmarczyk2017energy} to incorporate the influence of spatially varying bone density. Furthermore, it combines this with bone adaptation \citep{kaczmarczyk2011efficient,lewandowski2017}. 

Although this work is generic in nature and applicable to both human and animal bone, this paper focuses the numerical examples on equine bone. 

To the best of the authors' knowledge, to date there is only one report of equine bone adaptation in a FEM framework~\citep{Wang2016}. 
In that work, a mechanostat micro-scale model of three-dimensional cortical bone remodelling, informed by \emph{in~vivo} equine data, was presented. 
The model used the von Mises stress as a stimulus to control microstructural cortical bone remodelling.
In contrast, the current paper presents a full macro-scale model of equine bone response to mechanical loading, testing a hypothesis that micro-damage and fracture can be modelled at the macroscale by using clinically available CT-scanning data.
The motivation for this work is to generate subject-specific simulations to acquire meaningful insight into bone resistance
 for veterinary practitioners.

This article is structured as follows. After establishing the kinematic preliminaries in Section~\ref{preliminaries}, the mathematical framework for bone adaptation is briefly presented in Section~\ref{sec:bone_remodel}. Section~\ref{sec:fracture} extends the authors' previous work for evolving crack propagation in the context of configurational mechanics. The method is utilised to calculate fracture resistance and crack propagation under quasi-static loading during different stages of adaptation. Section \ref{sec:fem_modelling} describes the finite element method implementation and Section~\ref{sec:fem_modelling} describes a special element for capturing the singular stress field at the crack front.  
All the above components are brought together into a single framework and its performance is demonstrated using a series of numerical examples in Section~\ref{sec:numerical_examples}.

 \section{Preliminaries}
 \label{preliminaries}
\begin{figure}[th] 
\setlength{\fboxsep}{0pt}%
\setlength{\fboxrule}{0pt}%
\begin{center}
\includegraphics[width=10cm]{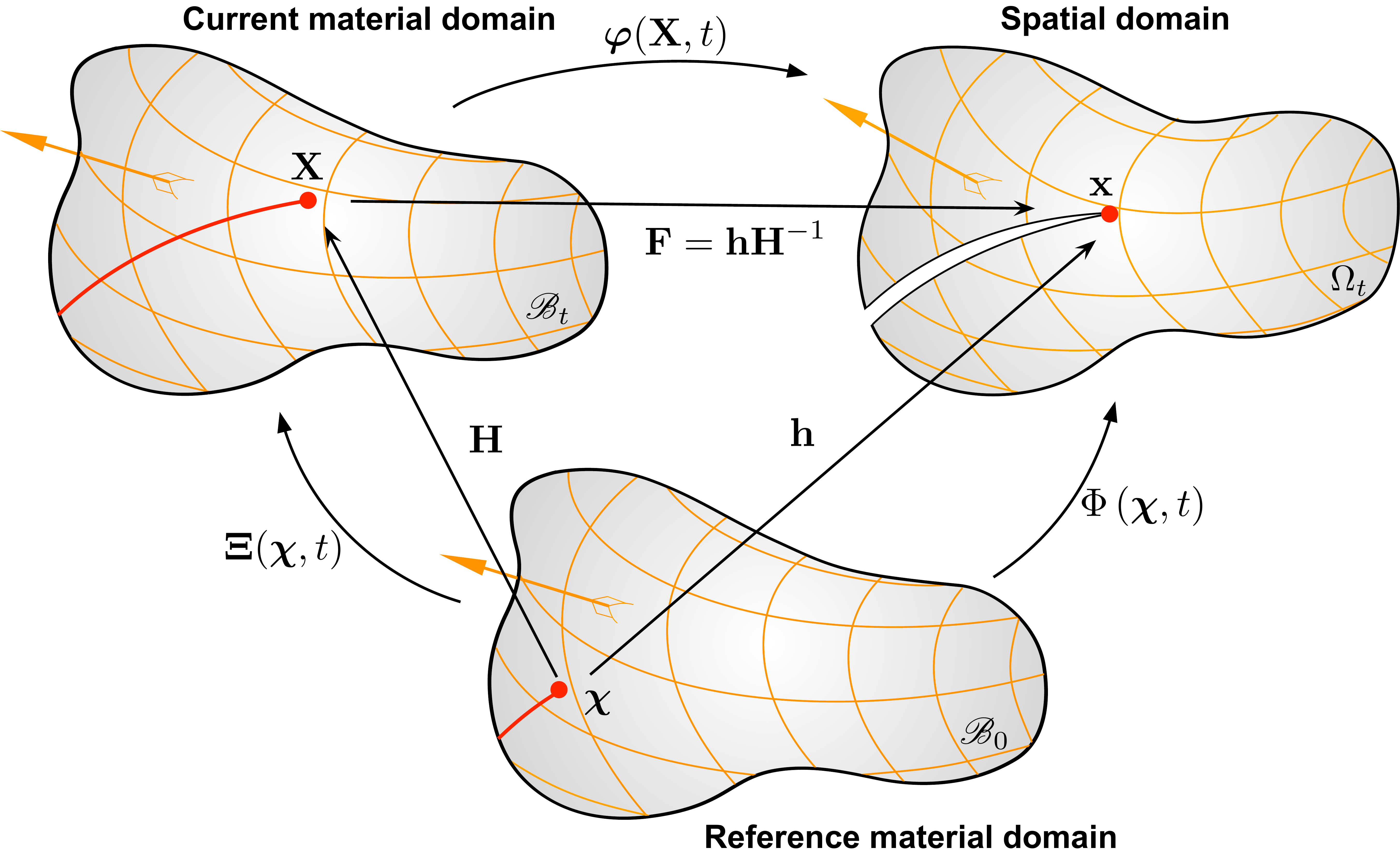} 
\end{center}
\caption{Kinematics of crack propagation in elastically deforming bone.}
\label{fig:domains4}
\end{figure}

Figure~\ref{fig:domains4} shows a section of bone with an initial crack in the reference domain $\mathscr{B}_{0}$. As a result of loading, the crack extends and the body deforms elastically. Working within the framework of configurational mechanics \citep{kaczmarczyk2014three,kienzler2014configurational}, it is convenient to decompose this behaviour into an extension of the crack in the material domain  $\mathscr{B}_t$ followed by elastic deformation  in the spatial domain $\Omega_t$. The former is described by the mapping from the reference  domain to the material domain ${\boldsymbol\Xi}$, whilst the latter is described by the mapping from the material to the spatial domain ${\boldsymbol\varphi}$ - Figure~\ref{fig:domains4}.

The material coordinates $\mathbf{X}$ are mapped onto the spatial coordinates $\mathbf{x}$ via the
familiar deformation map $\boldsymbol\varphi(\mathbf{X},t)$. The physical displacement is:
\begin{equation}
\mathbf{u}=\mathbf{x}-\mathbf{X}
\end{equation}
The reference material domain describes the body before crack extension. ${\boldsymbol\Xi}(\boldsymbol\chi,t)$ maps the reference material coordinates $\boldsymbol\chi$ on to the current material coordinates $\mathbf{X}$, representing a configurational change, i.e. extension of the crack due to advancement of the crack front. ${\boldsymbol\Phi}$ maps the reference material coordinates $\boldsymbol\chi$ on to the spatial coordinates $\mathbf{x}$. The current material and spatial displacement fields are given as
\begin{equation}
\mathbf{W} = \mathbf{X} - {\boldsymbol\chi}\quad\textrm{and}\quad
\mathbf{w} = \mathbf{x} - {\boldsymbol\chi}
\end{equation}
$\mathbf{H}$ and $\mathbf{h}$ are the gradients of the material and spatial maps and $\mathbf{F}$ the deformation gradient~\cite{kaczmarczyk2014three}, defined as:
\begin{equation}
\mathbf{H}=\frac{\partial {\boldsymbol\Xi}}{\partial {\boldsymbol\chi}},\quad\mathbf{h}=\frac{\partial {\boldsymbol\Phi}}{\partial {\boldsymbol\chi}},\quad\mathbf{F} = \frac{\partial \boldsymbol\varphi}{\partial \mathbf{X}} = \mathbf{h}\mathbf{H}^{-1}
\end{equation}

The time derivative of the physical displacement $\mathbf{u}$ and the deformation gradient $\mathbf{F}$ (material time derivative) are given as~\cite{kaczmarczyk2014three}:
\begin{equation}\label{eq:phy_vel}
\dot{\mathbf{u}}= \dot{\mathbf{w}}-\mathbf{F}\dot{\mathbf{W}} \qquad 
\dot{\mathbf{F}} = \nabla_\mathbf{X} \dot{\mathbf{x}} = \nabla_\mathbf{X} \dot{\mathbf{u}} = 
\nabla_\mathbf{X} \dot{\mathbf{w}} - \mathbf{F} \nabla_\mathbf{X} \dot{\mathbf{W}}
\end{equation}
 
\section{Bone adaptation} \label{sec:bone_remodel}
In this paper, the modelling of bone adaptation is based on the work of Kuhl and Steinmann~\citep{kuhl2003theory} in which bone is considered an elastic porous material. The model is stable~\citep{kuhl2003computational}, efficient~\citep{kaczmarczyk2011efficient} and capable of 
producing bone mineral density profiles that are quantitatively comparable with DEXA scans following gait analysis~\citep{pang2012computational}. 
Using this approach, bone adaptation in human scapula~\citep{liedtke2017computational}, 
tibia~\citep{pang2012computational}, humerus~\citep{taylor2009phenomenon} and femur with various surgical implants~\citep{ambrosi2011perspectives, Connor2017bone} have been simulated and its potential in topology optimization~\citep{waffenschmidt2012application} has been explored. 
One of the advantages of such a phenomenological approach is that only a small number of parameters are required, which can be experimentally
determined from, for example, CT imaging~\citep{zadpoor2013open}.

 \subsection{Conservation of mass}

Following Kuhl and Steinmann~\citep{kuhl2003computational}, it is assumed that
the rate of change of the time-dependent material density is in equilibrium
with mass flux, expressed as:
\begin{equation} \label{eq:mass_balance}
\frac{\partial\rho}{\partial t}  + \dot{\mathbf{W}} \cdot  \nabla_\mathbf{X} \rho = 
\nabla_{\boldsymbol {\rm X}} \cdot \mathbf{R} + \mathcal{R}_0
\end{equation}
where $\rho$ is mass density and $\mathcal{R}_0$ is the locally created mass.
Furthermore, $\mathbf{R}$ is the mass flux defined as:

\begin{equation}
\mathbf{R} = \mathcal{R} \nabla_\mathbf{X} \rho
\label{eq:mass_flux}
\end{equation}

where $\mathcal{R}$ is mass conductivity. The term on the right hand side of Eq. \ref{eq:mass_balance}
is the material time derivative associated with the evolving current material
configuration. 
However, in the present work it is assumed that the mass flux $\mathbf{R}$ is zero and hence only the local mass source $\mathcal{R}_0$ contributes to the changes in density.

\subsubsection{Constitutive relationship for bone adaptation}

\label{sec:constitutive_eq}

\begin{figure}[!htb]
	\centering
		\input{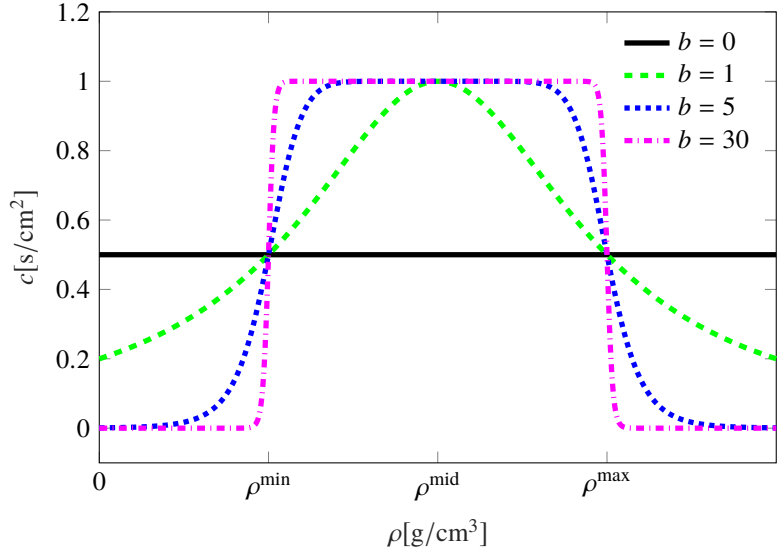}
		\caption{Bell function for parameter $c$ for different values of the integer exponent $b$. 
		As $b \rightarrow \infty$, the bell-shape curve becomes infinitely steep at 
		$\rho{^\mathrm{min}}$ and $ \rho{^\mathrm{max}}$.}
		\label{fig:bell_func}
\end{figure}

Following Harrigan and Hamilton~\citep{Harrigan1993}, the constitutive
relation for the mass source is:
\begin{equation}
\mathcal{R}_{0}=c\left[\Biggl[\frac{\rho}{\rho_{0}^{\ast}}\Biggr]^{-m}\Psi
-\Psi^{\ast}\right]
\label{eq:mass_source}
\end{equation}
where $\rho_0^\ast$ and $\Psi^\ast$ represent reference values of the
density, $\rho$ and free energy, $\Psi$, respectively. The driving term
$\left[ \rho / \rho_0^\ast \right]^{-m}\Psi$ tends to converge to
$\Psi^\ast$ (see Eq.~(\ref{eq:mass_source})) when density saturation is
achieved and local generation of bone ceases. The exponent $m$ is a
dimensionless scalar introduced to guarantee uniqueness and
stability~\citep{Harrigan1993} . The coefficient $c$ controls
the rate of the adaptation process with units~$[\rm{s/cm^2}]$. As proposed
in~\citep{Waffenschmidt2012}, it can be beneficial to prescribe an upper and
lower bound for bone density, thereby avoiding spurious or non-physical
values. In this paper, the parameter $c$, which is conventionally considered 
to be constant, is replaced by a bell function
defined as:
\begin{equation}
\begin{aligned}
c(\rho) = & \frac{1}{1 + \left[  (\rho - \rho^{\mathrm{mid}}) / 
(\rho{^\mathrm{max}} - \rho{^\mathrm{mid})} \right]^{2 b}}\\
& \mathrm{with} \quad \rho^{\mathrm{mid}} = 
\frac{\rho{^\mathrm{max}} + \rho{^\mathrm{min}}}{2}
\end{aligned}
\label{eq:bell_function}
\end{equation}
$\rho^\mathrm{max}$ and $ \rho{^\mathrm{min}}$ where $\rho^\mathrm{max}$
and $\rho^\mathrm{min}$ are the maximum and minimum values of $\rho$, and
$\rho^{\rm {mid}}$ is their average. The bell function
(\ref{eq:bell_function}) is illustrated in Figure~\ref{fig:bell_func} for
different values of the integer exponent, $b$. Its application and influence
on the overall results are elaborated in
Section~\ref{sec:numerical_examples}.

\subsection{Elastic constitutive relationship}
As an elastic porous material, the free energy $\Psi (\mathbf F, \rho)$, for bone is taken as
\begin{equation}
\Psi=\left[\frac{\rho}{\rho_{0}^{\ast}}\right]^{n}\Psi^{\mathrm{neo}},
\label{eq:free_energ}
\end{equation}
where $\Psi^{\rm {neo}}$ is the Helmholtz free energy for a Neo-hookean material, which is expressed in terms of the right
Cauchy-Green deformation tensor $\boldsymbol{\rm{C}}$:
\begin{equation}
\Psi^{\mathrm{neo}}=\frac{\mu}{2}\left[\textrm{tr}(\mathbf{C})-3\right]-\mu\ln(\sqrt{\det\mathbf{C}})+\frac{\lambda}{2}\ln^{2}(\sqrt{\det\mathbf{C}})
\end{equation}
where $\mu$ and $\nu$ are the Lam\'e constants. Moreover, the exponent $n$ is a
non-physical parameter that typically varies as $1 \leq n \leq 3.5$, 
depending on the porosity of the material~\citep{Gibson2005}. 

Bone adaptation is a mechanically driven process, whereby the density field evolves in response to the mechanical environment. Likewise, the material stiffness is directly dependent on the density and this, in turn, influences the mechanical response. Therefore, the equation for conservation of mass \ref{eq:mass_balance} is coupled with the equation for linear momentum balance:
\begin{equation} \label{eq:linear_momentum}
\nabla_{\mathbf X} \cdot \mathbf P = 0
\end{equation}
where $\mathbf P$ is the the first Piola-Kirchhoff stress:  
\begin{equation}
\mathbf P = \frac{\partial \Psi (\mathbf F, \rho)}{\partial \mathbf F}
\end{equation}

This coupled system of equations is solved using the finite element method - see Section \ref{sec:numerical_examples:bone_adap}.

\section{Fracture resistance and fracture propagation}
\label{sec:fracture}
Various theories exist in the literature regarding failure criteria for bone tissue and it is now common practice for researchers to estimate fracture resistance within the framework of FEM. In particular, subject-specific FEM models can potentially quantify the risk of failure under a given loading scenario. However, this still remains an open challenge.

In recent years, the main focus in bone mechanics was in the use of different strength criteria for the onset of failure. 
The most commonly adopted ones were based on stress~\citep{keyak2005predicting} or strain measures~\citep{schileo2008subject} assuming bone failure is determined by a yield criterion~\citep{yosibash2010predicting}. 
Experimental validation of such simplified models show that there is a significant spread in the predicted failure, with errors between 10\% and 20\%~\citep{van2014accurately}.
This variation is explained perhaps by the focus on the local initiation of failure, rather than the complete failure mechanism. 
The fracture process of bone is very important,  particularly in the case of fatigue fractures~\citep{gupta2008fracture}. 
Limitations in previous studies (e.g. use of 2D geometry~\citep{bettamer2017using}, assuming homogeneous bone properties~\citep{gasser2007numerical}), 
can also explain why an appropriate model for bone fracture has not been developed previously.

This paper builds on the authors' computational framework~\citep{kaczmarczyk2017energy} for brittle fracture within the context of configurational mechanics, extending it here to include the influence of heterogeneous materials such as bone.
The concept of configurational forces was originally introduced by Eshelby~\citep{eshelby1951force}. 
Unlike physical forces, configurational forces act on the material manifold and represent the tendency of imperfections like cracks, voids or material inhomogeneities to move relative to the surrounding material. 
The past two decades have seen a growing interest in this approach for analysis of material imperfections~\citep{maugin2016configurational} and in particular for evaluating the forces driving crack advancement~\citep{kaczmarczyk2017energy,steinmann2001application, ozencc2016configurational}. 
However, until recently this approach has never been used to effectively assess configurational forces in heterogeneous bodies with cracks. 

In the current study, additional configurational forces arising from inhomogeneities~\citep{kienzler2014configurational}
associated with spatially varying bone density are introduced into the formulation. This allows for the accurate assessment of the likelihood of a crack to propagate and to simulate  the subsequent propagation of fractures in bone. 
An additional goal is to investigate bone fracture at different stages of bone adaptation, utilising either the results from bone adaptation analysis (Section~\ref{sec:bone_remodel}) or data directly taken from CT scans. 
Similar concepts of combined adaptation and fracture analyses has been presented before~\citep{hambli2013integrated}. 
However, it utilised a different adaptation model and continuum damage mechanics approach for fracture, both of which require many more parameters to calibrate. 

\subsection{First and second laws of thermodynamics}

\begin{figure}[th]
\setlength{\fboxsep}{0pt}%
\setlength{\fboxrule}{0pt}%
\begin{center}
\def\svgwidth{12cm} 	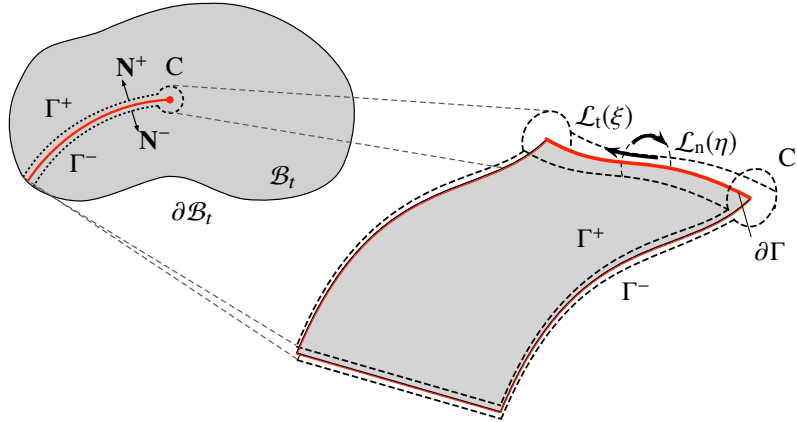 
\end{center}
\caption{Crack construction. In 2D (left) and in more detail in 3D (right).}
\label{fig:crac_surf_construct} 
\end{figure}

The first law of thermodynamics can be expressed as 
\begin{equation}
\label{eq::first_law}
\int_{\partial \mathcal B_t} {\dot {\boldsymbol {\rm u} } } 
\cdot \mathbf t \mathrm d S = \int_{\partial \Gamma } \gamma \dot A_\Gamma +
{\frac{\mathrm d}{\mathrm d t }} 
\int_{\mathcal B_t} \Psi(\mathbf F, \rho) \mathrm d V
\end{equation}
where the left hand side is the power of external work, the first term on the
right hand side is the rate of crack surface energy and the last term is
the rate of internal energy. $\mathbf t$ is the external
traction vector, $\gamma $ is the surface energy $[ {\rm{N m}}^{-1} ]$, $\dot{A}_\Gamma$ is the change in the
crack surface area and
$\Psi$ is the volume specific free energy. The crack surface $\Gamma$ comprises two crack faces and a crack front $\partial\Gamma$ - see Figure~\ref{fig:crac_surf_construct}.

In~\citep{kaczmarczyk2017energy}, a kinematic relationship between the change in the
crack surface area $\dot{A}_\Gamma$ and the crack front velocity $\dot{\mathbf{W}}$ was derived that is given as:
\begin{equation}
\label{eq::Agamma2}
\dot{A}_\Gamma
 =
\int_{\partial\Gamma}
\mathbf{A}_{\partial\Gamma} \cdot \dot{\mathbf{W}} \textrm{d}L
\end{equation}
where 
$\mathbf{A}_{\partial\Gamma}$ is a dimensionless kinematic state variable that defines the orientation of the current crack front that can be considered a unit vector normal to the crack front and tangential to the crack surface. In deriving this expression, it was recognised that any change in the crack surface area $\dot{A}_\Gamma$ in the current material space can only occur due to motion of the crack front.

Making use of Equations~(\ref{eq:phy_vel}) and (\ref{eq::Agamma2}),
and given that $\mathrm d \dot V = \nabla _{\mathbf X} \cdot \mathbf{\dot W}
\mathrm d V$, Eq~(\ref{eq::first_law}) can be reformulated as:

\begin{equation}\label{eq:crack_first_law}
\int_{\partial \mathcal B_t} \big( {\dot {\boldsymbol {\rm w} } } \cdot \mathbf t - 
{\dot {\boldsymbol {\rm W} } } \cdot \mathbf F^{\rm T} \mathbf t \big) \mathrm d S = 
\int_{\partial \Gamma } \gamma \mathbf A_{\partial \Gamma} 
\cdot {\dot {\boldsymbol {\rm W} } } \mathrm d L + \int_{\mathcal B_t} 
\big( \mathbf P\, \colon \nabla_{\mathbf X} {\dot {\boldsymbol {\rm w} } } + 
{\bm  \Sigma}\, \colon \nabla_{\mathbf X} {\dot {\boldsymbol {\rm W} } } 
+  \mathbf f^{\mathrm {inh}} \cdot \dot{\mathbf{W}}
\big) \mathrm d V 
\end{equation}
where
\begin{equation}
{\bm {\Sigma}} = \Psi(\mathbf F, \rho) \mathbf  1 - \mathbf F^{\rm T} 
\mathbf P(\mathbf F, \rho),
\, \mathrm{and} \quad
\mathbf f^{\mathrm {inh}} = 
\left.
\frac{\partial \Psi}{\partial \rho}
\right|_{(\mathbf{F} = {\rm{const}})}
\frac{\partial \rho}{\partial \mathbf X} 
\end{equation}
${\bm {\Sigma}}$ is the Eshelby stress tensor and
$\mathbf f^{\mathrm {inh}}$ is an additional fictitious force that arises from variations in the density field and drives the crack front from dense to less dense material. 

The spatial conservation law of linear momentum balance is repeated here:
\begin{equation} \label{eq:linear_momentum2}
\nabla_{\mathbf X} \cdot \mathbf P = 0
\;
\forall \mathbf{X}\in\mathcal B_t,
\quad
\mathbf{P}\mathbf{N} = \mathbf{t}\;
\forall \mathbf{X}\in\partial\mathcal B_t^\sigma
\end{equation}
where $\partial\mathcal B_t^\sigma$ is the region of the boundary where tractions are applied. 

The equivalent material momentum balance is expressed as:
\begin{equation}
\nabla_{\mathbf X } \cdot {\bm {\Sigma}}= \mathbf f^{\mathrm {inh}}
\;
\forall \mathbf{X}\in\mathcal B_t,
\quad
{\bm {\Sigma}}\mathbf{N} = \mathbf{F}^\textrm{T}\mathbf{t}\;
\forall \mathbf{X}\in\partial\mathcal B_t^\sigma
\end{equation}
It is important to note that $\mathbf f^{\mathrm {inh}}=\mathbf{0}$ in the case of homogeneous materials, with uniform density distribution.

After applying the divergence theorem to Eq.~(\ref{eq:crack_first_law}) and recognising the momentum balance laws, we follow~\citep{kaczmarczyk2017energy} to establish a local form of Eq.~(\ref{eq:crack_first_law}), which represents an expression for equilibrium of the crack front as
\begin{equation}\label{eq:crack_local_first_law}
	{\dot {\boldsymbol {\rm W} } } \cdot 
	\left( \gamma \mathbf A_{\partial \Gamma} - \mathbf G \right) = 0
\end{equation}
where the configurational force $\mathbf{G}$ is the driving force for crack propagation:  
\begin{equation}
	\mathbf G = \lim_{|\mathcal{ L }|\to 0} 
	\int_{ {{\mathcal L}_{\rm n}} } {\bm {\Sigma}}\mathbf{N}\, \mathrm d L 
	\label{eq:crack_configuration_force}
\end{equation}
From this equation, it is clear that the crack front is in equilibrium when the crack is not propagating, i.e. material
velocity $\dot{\mathbf{W}}$ at the crack front is zero, or when the crack front is propagating and the configurational force
is in equilibrium with the material resistance $\gamma \mathbf A_{\partial \Gamma}$. 

It should be noted that crack front equilibrium is unaffected by material heterogeneities and does not depend on $\mathbf f^{\mathrm {inh}}$. All terms in Eq.~\ref{eq:crack_local_first_law} are only evaluated at the crack front. However, it will be shown in Section~\ref{sec:fem_fracture_prop} that, in a discrete setting, calculation of the nodal configurational forces involves a volume integral of the density gradient.

Since Eq.~(\ref{eq:crack_local_first_law}) has more than one solution at equilibrium, depending on whether the crack 
does or does not propagate, the formulation is supplemented by a straightforward criterion for crack growth, equivalent to Griffith's  criterion~\citep{kaczmarczyk2017energy}:
\begin{equation} \label{eq:grif1}
\phi(\mathbf{G}) = 
\mathbf{G} \cdot \mathbf{A}_{\partial\Gamma} - g_c/2 \leq 0
\end{equation} 
where $g_c=2\gamma$ is a material parameter specifying the critical threshold of energy release
per unit area of the crack surface $\Gamma$, also known as the Griffith energy. For a point on the crack front to satisfy the crack growth criterion, 
either $\phi<0$ and $\dot{\mathbf{W}}=0$, or $\phi=0$, $\dot{\mathbf{W}}\ne 0$ and $\gamma\mathbf{A}_{\partial\Gamma}=\mathbf{G}$.

The direction of fracture propagation is
constrained by the second law of thermodynamics. Here we assume that fracture takes place
relatively fast compared to the process of adaptation (with no healing), 
such that non-negative dissipation at the crack front can be expressed as
\begin{equation}
	\mathcal{D} = \gamma \dot{\mathbf{W}} \cdot \mathbf{A}_{\partial\Gamma}= \dot{\mathbf{W}} \cdot \mathbf{G}\ge0
\end{equation}

It should be noted that the well-established stress intensity factors are not applicable 
in the case of heterogeneous materials, since it requires the existence of
an analytical solution for the stress field in the vicinity of the crack front that is
independent of arbitrary distribution of density. 
Similarly, the use of
J-integrals requires integration over the closed surface without
inhomogeneities (including heterogeneous density distribution), except for the crack front itself and therefore not applicable in this case.
Finally, it is worth noting that the current framework is formulated within 
the realm of large displacements and large strains, hence it 
is generally valid under any assumption for strains and displacements.

\subsection{Density field}
\label{sec:dens_mapping}
The previous subsections have shown that  fracture modelling of bone is influenced by the density distribution in the material configuration. This density field can be generated from either (a) a bone adaptation analysis, solving both Eqs~(\ref{eq:mass_balance}) and (\ref{eq:linear_momentum2}), or (b) subject-specific data (geometry and material properties) available from, for example, computed tomography (CT) scans. Previous examples in the literature of subject-specific modelling to assess the stresses and fracture resistance of bones can be found in~\citep{poelert2013patient,Helgason2008b,Yosibash2010}. Most algorithms that use voxel data have simply averaged~\citep{zannoni1999material} or integrated data onto finite elements, thereby supplying a constant density within their volume~\citep{taddei2007material, schileo2008subject}. In this paper, radiopacity associated with each 3D voxel from CT scan data is spatially approximated. 

In the numerical examples described later, both sources of density data are used. It will be shown in the next section that, in order to evaluate the configurational forces at the crack front, it is necessary to have a spatially smooth density field. Therefore, discrete density data will need to be approximated as a smooth density field, and this will be achieved by adopting the Moving Weighted Least Squares (MWLS) method. This mapping approach was chosen since it offers higher regularity (i.e. higher derivatives exist) than when the field is
directly approximated on the finite element mesh. Full details are given in \citep{karol_lewandowski_moving_2019}.  

\section{Finite element modelling} \label{sec:fem_modelling}

This section considers the sequential analysis of bone adaptation and fracture propagation, although it is recognised that the density field could be obtained directly from subject-specific data, in which case it may not be necessary to undertake the bone adaptation analysis. A sequential approach is justified since the process of bone adaptation takes place at a much longer time scales than fracture.

Three-dimensional domains are discretised with tetrahedral finite elements. 
Fields are approximated in the current material and current spatial spaces with 
hierarchical basis functions of arbitrary polynomial order, following the work of Ainsworth and Coyle~\citep{Ainsworth2003}.  
\begin{eqnarray}
	\rho^h({\boldsymbol {\rm \upchi}},t) = \pmb\Phi({\boldsymbol {\rm {\upchi}}}) {\tilde{\pmb\uprho}}(t) \\
	\mathbf{ X}^h({\boldsymbol {\rm \upchi}},t) = \pmb\Phi({\boldsymbol {\rm {\upchi}}}) \tilde {\mathbf{ X}}(t), 
	\quad \mathbf{x}^h({\boldsymbol {\rm \upchi}},t) = \pmb\Phi({\boldsymbol {\upchi}}) {\tilde{\mathbf{x}}}(t) \\
	\mathbf{ W}^h({\boldsymbol {\rm \upchi}},t) = \pmb\Phi({\boldsymbol {\rm \upchi}}) {\dot { \tilde {\boldsymbol {\rm W} } }}(t), 
	\quad \mathbf{w}^h({\boldsymbol {\rm \upchi}},t) = \pmb\Phi({\boldsymbol {\upchi}}){\dot { \tilde {\boldsymbol {\rm w} } }}(t)
	\label{eq:discretisation}
\end{eqnarray}
where $\mathbf{\Phi}$ are shape functions, superscript $h$ indicates approximation and $(\tilde \cdot)$ nodal
values. Moreover, the smoothed density field is approximated by MWLS shape functions
\begin{equation}
	\rho^{h,\textrm{MWLS}}(\mathbf{X},t) = \Phi^\textrm{MWLS}(\mathbf{X}) 
	\tilde{\pmb\uprho}^h(\Xi({\boldsymbol {\rm \upchi}}),t) 
\end{equation}
It should be noted that shape functions $\Phi^\textrm{MWLS}(\mathbf{X})$ are evaluated at
current material points, $\mathbf{X}$, rather than reference points, $\pmb\upchi$, as presented in Eq.~(\ref{eq:discretisation}) with the property of partition of unity.
Since the density field is evaluated at $\mathbf{X}$, the approximation is independent of changes of the material configuration (i.e.
changing mesh).

\subsection{Bone adaptation}
The bone adaptation problem is solved with a staggered approach, the material configuration is fixed
such that:
\begin{equation}
	\tilde {\mathbf{ X}}(t)	= \tilde {{\boldsymbol {\rm \upchi}}} =  \textrm{const}
	\;\;\textrm{and}\;\;
	\dot{\tilde {\mathbf{X}}}(t) = \dot{\tilde {\mathbf{W}}}(t) = 0
\end{equation}
where $\tilde {{\boldsymbol {\rm \upchi}}}$ is vector of nodal positions. The semi-discrete
form of equations (\ref{eq:mass_balance}) and (\ref{eq:linear_momentum}) take
the form of residuals
\begin{equation}
	\left\{
	\begin{split}
		\mathbf{r}^\rho(\tilde{\pmb\uprho}(t), \tilde{\mathbf{x}}(t)) =
		\int_{\mathcal{B}^h_t} \pmb{\Phi}^\textrm{T}\dot{\tilde{\pmb\uprho}}^h({\boldsymbol {\rm \upchi}},t) 
		\textrm{d}V 
		+
		\int_{\mathcal{B}^h_t} \nabla_\mathbf{X} \pmb{\Phi}^\textrm{T} 
		\mathcal{R} \nabla_\mathbf{X}\pmb{\Phi} \tilde{\pmb\uprho}
		\textrm{d}V
		-	
		\int_{\mathcal{B}^h_t} \pmb{\Phi}^\textrm{T} \mathcal{R}^h_0 
		\textrm{d}V	
		-
		\int_{\partial\mathcal{B}^h_t} \pmb{\Phi}^\textrm{T} 
		\mathbf{q}^\textrm{external} 
		\textrm{d}S	
		= \mathbf{0}\\
		\mathbf{r}^x(\tilde{\pmb\uprho}(t), \tilde{\mathbf{x}}(t)) =
		\int_{\mathcal{B}^h_t} \nabla_\mathbf{X}\pmb{\Phi}^\textrm{T}\mathbf{P}^h\textrm{d}V
		-
		\int_{\partial\mathcal{B}^h_t} \pmb{\Phi}^\textrm{T}\mathbf{f}^\textrm{ext, adapt}\textrm{d}S	
		= \mathbf{0}
	\end{split}
	\right.
\end{equation}
where $\mathbf{r}^\rho$ is the vector of residuals related to mass density flux 
equilibrium, $\mathbf{q}^\textrm{external}$ is influx of mass across the boundary, 
$\mathbf{r}^x$ is the vector of residuals associated with balance of linear momentum and
$\mathbf{f}^\textrm{ext, adapt}$ are averaged long term forces
mimicking mechanical load on the bone over long time period and $\mathcal{R}$ is mass conductivity. A truncated Taylor series expansion leads to the  semi-discrete form, expressed as:
\begin{equation}
	\left[
		\begin{array}{c}
			\mathbf{r}^\rho(\tilde{\pmb\uprho}_i(t), \tilde{\mathbf{x}}_i(t))\\
			\mathbf{r}^x(\tilde{\pmb\uprho}_i(t), \tilde{\mathbf{x}}_i(t))
		\end{array}
	\right]
	+
	\left[
	\begin{array}{cc}
		\mathbf{M}_{\rho\rho} & \mathbf{0}\\
		\mathbf{0} & \mathbf{0}
	\end{array}
	\right]
	\left\{
		\begin{array}{c}
		\delta\dot{\tilde{\pmb\uprho}}_{i+1}(t)\\
		\delta\dot{\tilde{\mathbf{x}}}_{i+1}(t)
		\end{array}
	\right\}
	+
	\left[
		\begin{array}{cc}
			\mathbf{K}_{\rho\rho} & \mathbf{K}_{\rho x}\\
			\mathbf{K}_{x\rho} & \mathbf{K}_{xx}
		\end{array}
		\right]
		\left\{
			\begin{array}{c}
			\delta\tilde{\pmb\uprho}_{i+1}(t)\\
			\delta\tilde{\mathbf{x}}_{i+1}(t)
			\end{array}
		\right\}
		=
		\left[
			\begin{array}{c}
				\mathbf{0}\\
				\mathbf{0}
			\end{array}
		\right]
		\label{eq:tangent_remodelling}
\end{equation}
with
\begin{equation}
\begin{split}
\mathbf{M}_{\rho\rho} = 
\left.
\frac{\partial \mathbf{r}^\rho}{\partial \dot{\tilde{\pmb\uprho}}}
\right|_{(\tilde{\pmb\uprho}_i(t),\tilde{\mathbf{x}}_i(t))},\,
\mathbf{K}_{\rho\rho} = 
\left.
\frac{\partial \mathbf{r}^\rho}{\partial {\tilde{\pmb\uprho}}}
\right|_{(\tilde{\pmb\uprho}_i(t),\tilde{\mathbf{x}}_i(t))},\,
\mathbf{K}_{\rho x} = 
\left.
\frac{\partial \mathbf{r}^\rho}{\partial {\tilde{\mathbf{x}}}}
\right|_{(\tilde{\pmb\uprho}_i(t),\tilde{\mathbf{x}}_i(t))},\,\\
\mathbf{K}_{x \rho} = 
\left.
\frac{\partial \mathbf{r}^x}{\partial {\tilde{\pmb\uprho}}}
\right|_{(\tilde{\pmb\uprho}_i(t),\tilde{\mathbf{x}}_i(t))},\,
\mathbf{K}_{x x} = 
\left.
\frac{\partial \mathbf{r}^x}{\partial {\tilde{\mathbf{x}}}}
\right|_{(\tilde{\pmb\uprho}_i(t),\tilde{\mathbf{x}}_i(t))},
\end{split}
\end{equation}
and 
\begin{equation}
\tilde{\pmb\uprho}_{i+i}(t) =
\tilde{\pmb\uprho}_{i}(t) + \delta\tilde{\pmb\uprho}_{i+i}(t),\;
	\tilde{\mathbf{x}}_{i+i}(t) =
	\tilde{\mathbf{x}}_{i}(t) + \delta\tilde{\mathbf{x}}_{i+i}(t)
\end{equation}
where $(\cdot)_i$ is quantity at Newton iteration $i$. Finally, the
above semi-discrete problem is discretised in time using implicit Euler scheme:
\begin{equation}
	\dot{\tilde{\pmb\uprho}}_{i+i}^{n+1}(t) =
	\frac{
	\tilde{\pmb\uprho}_{i+i}^{n+1}-\tilde{\pmb\uprho}^{n}
	}{\Delta t}
\end{equation}
where $\Delta t$ is length of time step, and $n$ is time step number.

Note that the density field variables are approximated using polynomial bases functions that are one order less than those used for the spatial position variables, thereby ensuring a stable solution without oscillations.

The discretised balance equations are solved iteratively using the Newton-Raphson method for the displacements and density.

%
%
\subsection{Fracture propagation} \label{sec:fem_fracture_prop}

Given that the bone adaptation and fracture propagation problems have different boundary conditions and geometry they are solved as 
a staggered coupled problem. Initially bone adaptation is simulated under 
long-term effective loads applied without initial crack. Subsequently, an 
initial crack is inserted to compute the effect 
of short-term loads or extreme cyclic loading. The two different meshes for bone adaptation and fracture propagation 
are tailored for the specific analysis at hand. As
a consequence, the approximated Piola stress tensor for bone
adaptation is expressed as follows:
\begin{equation}
	\mathbf{P}^\textrm{h} = 
		\mathbf{P}(\mathbf{F}^\textrm{h}, \rho^h)
\end{equation}
whereas, for the fracture propagation problem, it is approximated as
\begin{equation}
	\mathbf{P}^{\textrm{h},\textrm{MWLS}} = 
		\mathbf{P}(
		\mathbf{F}^\textrm{h}, \rho^{h,\textrm{MWLS}})
\end{equation}

The residual force vector in the discretised spatial domain is expressed in the classical way as:
\begin{equation}
\label{spatial_residual}
	\mathbf{r}_\textrm{s}^\textrm{h}(\tilde{\pmb\uprho}(t), \tilde{\mathbf{x}}(t)) = \tau\mathbf{f}^\textrm{h}_\textrm{ext,s}-\mathbf{f}^\textrm{h}_\textrm{int,s}=\tau \int_{\partial\mathcal{B}^h_t} \pmb{\Phi}^\textrm{T}
	\mathbf{f}^\textrm{ext}
	\textrm{d}S-
	\int_{\mathcal{B}^h_t} \nabla_\mathbf{X} \pmb{\Phi}^\textrm{T}
	\mathbf{P}^{h,\textrm{MWLS}}\textrm{d}V=\mathbf{0}
\end{equation}
where $\tau$ is the unknown scalar load factor, $\mathbf{f}^\textrm{h}_\textrm{ext,s}$ is the vector of externally applied forces and $\mathbf{f}^\textrm{h}_\textrm{int,s}$ is the vector of internal forces. 

Discretisation of Eq.~\ref{eq:crack_local_first_law} establishes the material counterpart to Eq.~\ref{spatial_residual}, expressed as
\begin{equation}
\label{material_residual}
\mathbf{r}_\textrm{m}^\textrm{h}(\tilde{\pmb\uprho}(t), \tilde{\mathbf{x}}(t)) = \mathbf{f}^\textrm{h}_\textrm{res}-\tilde{\mathbf{G}}^\textrm{h}=\mathbf{0}
\end{equation}
$\tilde{\mathbf{G}}^\textrm{h}$ is the vector of nodal configurational forces only on nodes on the crack front, with the integration restricted to elements adjacent to the crack front:
\begin{equation}
	\tilde{\mathbf{G}}^\textrm{h} = 
	\int_{\mathcal{B}^h_t}
		\nabla_\mathbf{X}\pmb{\Phi}^\textrm{T} {\pmb\Sigma}^{h,\textrm{MWLS}}
	\textrm{d}V
	+
	\int_{\mathcal{B}^h_t}
		\pmb{\Phi}^\textrm{T} \dfrac{\partial {\Psi}^{h,\textrm{MWLS}} }{\partial \rho^{h,\rm{MWLS}}}
		\left(
			\frac{\partial 
			\rho^{h,\textrm{MWLS}}}{\partial \mathbf{X}}
		\right)
	\textrm{d}V
	\label{eq:mat_int_front}
\end{equation}
These configurational forces are the driving force for crack propagation. It should be noted that the second term of $\tilde{\mathbf{G}}^\textrm{h}$ reflects the influence of the spatially varying density. In the case of a homogeneous material, this second term would be zero. It should also be noted that this is only the case for the discretised configurational forces and that the continuum equivalent (Eq.~\ref{eq:crack_configuration_force}) is unaffected by variation in the density field.

$\mathbf{f}^\textrm{h}_\textrm{res}$ is the vector of nodal material resistance forces, given as:
\begin{equation}
\mathbf{f}^\textrm{h}_\textrm{res}=\frac{1}{2}\left(\tilde{\mathbf{A}}_\Gamma^\textrm{h}\right)^\textrm{T}\mathbf{g}_\textrm{c}
\end{equation}
where $\mathbf{g}_\textrm{c}=\mathbf{1}g_\textrm{c}$ is a vector of size equal to the number of nodes on the crack front. $\tilde{\mathbf{A}}_\Gamma^\textrm{h}$ defines the current orientation of the crack front and is a matrix comprising direction vectors along the crack front that are normal to the crack front and tangent to the crack surface:
\begin{equation}
\tilde{\mathbf{A}}_\Gamma^\textrm{h} = 
\int_{S^h_\Gamma}
\pmb{\Phi}^\textrm{T} 
\frac{\partial {A}^h_{\Gamma}}{
\partial \tilde{\mathbf{X}}}
\textrm{d}L
\end{equation}
$\tilde{\mathbf{A}}_\Gamma^\textrm{h}$ is evaluated by only integrating over $S_\Gamma^\textrm{h}$ that defines the area of those triangular finite elements that discretise the crack surface $\Gamma^\textrm{h}$ adjacent to the crack front $\partial\Gamma^\textrm{h}$.
$A^\textrm{h}_{\Gamma}$ is calculated as:
\begin{equation}
A^\textrm{h}_{\Gamma} = 	
	\| \mathbf{N}(\tilde{\mathbf{X}}) \|
=
\left\| 
\epsilon_{ijk}
\frac{\partial \Phi^\alpha_p}{\partial \xi_i}  
\frac{\partial \Phi^\beta_r}{\partial \xi_j} 
\tilde{X}^\alpha_p
\tilde{X}^\beta_r
\right\|
\end{equation}
where $\alpha, \beta \in \{0, \dots, N_{\rm{base}}\}$ are numbers of base functions,
$i,j,k,l,p,r \in \{0,1,2\}$ are material indices and $\boldsymbol{\upepsilon}$~is the Levi-Civita tensor. 
Moreover, the total number of degrees of freedom on element is $3(N_{\rm{base}}+1)$ and the units of $\tilde{\mathbf{A}}_\Gamma^\textrm{h}$ are $[{\rm m}^{-1}]$. $\mathbf{N}$ are the normals to the crack surface $\Gamma$.

\section{Singularity element}
\label{sec:singularity}
For the purposes of determining parameters such as stress intensity factors, it can be useful to reproduce the singular stress field at the crack front. However, conventional finite elements that adopt polynomial approximation functions are unable to do this. 
In this paper a new type of finite element with hierarchical approximation functions that overcome this problem is briefly presented. This is inspired by the so-called quarter-point elements, originally developed in the 1970s, whereby the mid-node of all edges connected to the crack tip node were shifted to the quarter-point~\citep{barsoum1976use,henshell1975crack}. In this work, all bodies are discretised using 3D tetrahedral elements. However, for simplicity, we present the main attributes in this paper in 1D.

For elements adjacent to the crack tip in the material configuration, the approximated material displacement field, using hierarchical shape functions (up to 2nd order), is expressed as:
\begin{equation}\label{eq:hierarchical_u}
W(\xi) = \sum_{a=0}^2 N_a (\xi) W^{(a)} = (1 -\xi)W^{(0)} + \xi W^{(1)} + \kappa (1 - \xi) \xi W^{(2)}
\end{equation}
where the natural coordinate $0\le \xi \le 1$ and $N_2 = \kappa N_0 N_1 =  \kappa\xi(1 - \xi)$. The parameter $\kappa$ is introduced, resulting in a nonlinear mapping between the natural and physical coordinates and leading to the desired singular stress and strain field at the crack tip. 

Adopting an isoparametric formulation, the element geometry can also be interpolated using the same approximation functions as for $W$. Thus, the physical distance from the crack tip is expressed as:
\begin{equation}\label{eq:hierarchical_r}
r_{\rm q}(\xi) = \sum_{a=0}^2 N_a (\xi) r_{\rm q}^{(a)} = \xi l + \kappa \xi(1-\xi)  l 
\end{equation}
where $r_{\rm q}(\xi=0)=0$ at the crack tip and $r_{\rm q}(\xi=1)=l$. Setting $\kappa=-1$ results in the following relationship:
\begin{equation}
r_{\rm q}= \xi l - \xi(1-\xi)l= \xi l \quad \Rightarrow \quad \xi = \sqrt{\frac{r_{\rm q}}{l}},
\end{equation}
This  yields the following radial dependence for displacements and strains:
\begin{equation}\label{eq:singstrain}
\begin{aligned}
W(r_{\rm q}) &= W^{(0)} + \left( -W^{(0)} + W^{(1)} - W^{(2)} \right) \sqrt \frac{r_{\rm q}}{l} - W^{(2)} \frac{r_{\rm q}}{l}\\
\varepsilon(r_{\rm q}) &= \frac{\partial W}{\partial r_{\rm q}} = \left( W^{(0)}  + W^{(1)} - W^{(2)}  \right) \frac{1}{2} \sqrt \frac{l}{r_{\rm q}} + W^{(2)} \frac{1}{l}
\end{aligned}
\end{equation}
Eq.~(\ref{eq:singstrain}) has the necessary terms to reproduce rigid body motion and pass the patch tests, as well as the desired singularity at the crack tip due the existence of the term $1 / \sqrt r_{\rm q}$. This will enable the elements adjacent to the crack front to reproduce the strain singularity resulting in an accurate finite element solution~\citep{nejati2015use}. 
The influence of this approach for tetrahedral elements will be investigated in Section~\ref{sec:release_energy_rate}.

\section{Numerical examples}
\label{sec:numerical_examples}
Several numerical examples are presented to illustrate each aspect of the proposed framework. 
The first set of analyses, presented in Subsection~\ref{sec:numerical_examples:bone_adap}, considers bone adaptation, using an equine 3rd metacarpal bone as a case study.
The performance of the singularity element formulation is demonstrated in Subsection~\ref{sec:release_energy_rate} using a finite plate with through thickness crack subjected to uniaxial stress. In the penultimate subsection, the framework is used to investigate the likelihood of fracture at different  stages of adaptation. The final example considers fracture propagation at different stages of adaptation.

\subsection{Bone adaptation examples}
\label{sec:numerical_examples:bone_adap}
This subsection considers the bone adaptation of an equine 3rd metacarpal bone. 
In the UK, approximately 60\% of horse fatalities at racecourses are directly or indirectly associated with a fracture, with the distal limb the most commonly affected site \citep{parkin2004risk}.
Most of these fractures occur due to the accumulation of tissue fatigue, as a result of repetitive loading~\citep{Parkin2005}, rather than a specific traumatic event. 
Intense exercise and excessive loading of the metacarpal bones results in maladaptation. 
The location of 3rd metacarpal fractures is remarkably consistent across a large number of racehorses,  with crack initiation presenting from 
the lateral para-sagittal groove of the distal condyle of the leading forelimb~\citep{jacklin2012frequency, parkin2006analysis}.
Despite considerable research in the field, including applying diagnostic methods such as radiography 
\citep{bogers2016quantitative, crijns2014intramodality, loughridge2017qualitative}, magnetic resonance imaging 
\citep{tranquille2017MRI} and biomarkers~\citep{mcilwraith2005use}, it still remains a challenge to accurately predict the fracture risk 
and prevent this type of significant injury.

\begin{figure}[h]
	\begin{center}
		   \begin{tikzpicture}
				\node at (0,0) {\includegraphics[width=8cm]{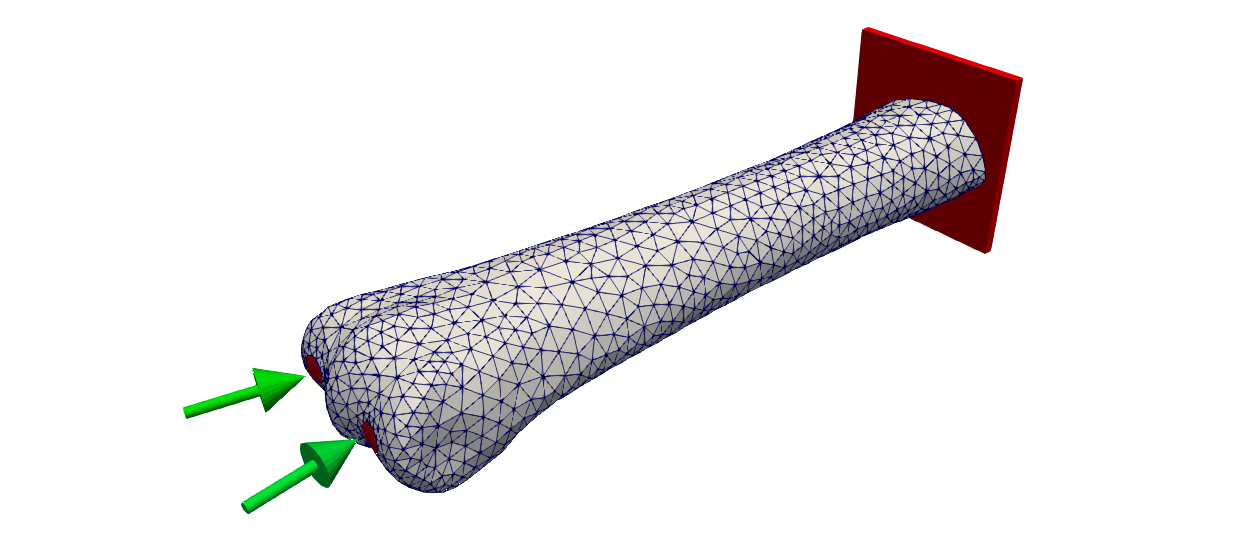}};
				\node at(-2.5,-1.8){$F = 5 \mathrm{kN}$ };
				\node at(-3.1,-0.3){ $F = 5 \mathrm{kN}$ };
				\node at(2.8,-0.3){ Fixed end };
	\end{tikzpicture}
		\caption{Finite element mesh of the equine 3rd metacarpal bone. The mesh consists of 14,041 quadratic tetrahedral elements and 70,901 degrees of freedom. To simulate the peak load of a gallop, 5 kN forces are applied on the lateral and medial side of the distal condyle.}
		\label{fig:mc3_BC}
	\end{center}
\end{figure}

Three cases are studied, using the material parameters presented in Table~\ref{tab:parameters_mc3}.
Stiffness and porosity values are derived from mechanical tests~\citep{Les1994}, whereas other values are from previous studies of human tibia~\citep{Pang2012,Waffenschmidt2012}. Each case considers a different function for the parameter $c$ that defines the rate of bone adaptation and is used to compute the mass source, $\mathcal{R}_0$, according to Eq.~(\ref{eq:mass_source}).
In Case~1, $c$ is constant. For Case~2 and Case~3 different bell functions (Eq.~(\ref{eq:bell_function})) are used. The parameters for each case are presented in Table~\ref{tab:three_cases}. 
The finite element mesh used in all cases comprises 17041 tetrahedral elements and was generated by discretising the segmented geometry of a full-scale model of an equine 3rd metacarpal bone derived from CT scan data - see Figure~\ref{fig:mc3_BC}.

For each analysis, the initial density is chosen to be homogeneous since, in the thermodynamic-based model, the starting density does not have a significant effect on the final bone density distribution (similar to other models at biological equilibrium~\citep{kuhl2003theory}). 
Boundary conditions are simplified to two representative forces (5~$\text{[kN]}$~each) spanning over a small area based on pressure film studies~\citep{Brama2001}, as demonstrated in Figure~\ref{fig:mc3_BC}. 
The two forces are often considered in the literature as an equivalent of joint peak force at the mid-stance of a horse gait. 
An adaptive time stepping scheme (using PETSc~\citep{petsc-web}) is used in all the simulations with an initial time step $\Delta t = 0.5$~$[\text{days (d)}]$, maximum time step of $\Delta t_{\text{max}} = 50$~$[\text d]$ and minimum of $\Delta t_{\text {min}} = 0.05$~$[\text d]$.
\begin{table}[h]
	\centering
	\begin{tabular}{lll}
		\hline
		Parameter             & Description                  & Value  \\ \hline
		$E  $                 & Young's modulus              & $4700 \,\mathrm{ [MPa]}$ ~\citep{Les1994} \\
		$\nu  $               & Poisson ratio                & $0.3 \,\mathrm{ [-]}$ \\
		$\rho_0 ^\ast  $      & Initial density              & $1.0 \,\mathrm{[ g/cm^{3}]}$  \\
		$\psi_{0}^\ast $      & Target energy density        & $0.0275\,\mathrm{ [MPa]}$  ~\citep{Waffenschmidt2012}  \\
		$c$                   & Density growth velocity      & $1.0 \,\mathrm{ [d/cm^{2}]}$   \\
		$m$                   & Algorithmic exponent         & $ 3.25 \,\mathrm{ [-]}$          \\
		$n$                   & Porosity exponent            & $2.25 \,\mathrm{ [-]}$     ~\citep{Les1994}   \\ 
		\hline
	\end{tabular} 
	\caption{Material parameters used for the simulations of 3rd metacarpal bone adaptation.}
	\label{tab:parameters_mc3}
\end{table}

\begin{table}[h]
	\centering
	\begin{tabular}{lllll}
		\hline
		Case                  & $c$              				       & $b$     &$\rho^\mathrm{max}$               &$\rho^\mathrm{min}$ \\ \hline
		1                     & 1              								 & -       & -                                & -\\
		2                     & Eq.~(\ref{eq:bell_function})   & 1000    & 2.5~$[{\text{g/cm}}^3]$          & 0.3~$[{\text{g/cm}}^3]$\\
		3                     & Eq.~(\ref{eq:bell_function})   & 30      & 1.8~$[{\text{g/cm}}^3]$          &1.0~$[{\text{g/cm}}^3]$ \\
		\hline
	\end{tabular} 
	\caption{Presentation of three cases input parameters for the evaluation of coefficient $c$ to compute mass source, $\mathcal{R}_0$, as presented in Eq.~(\ref{eq:mass_source}). All cases have common material input parameters presented in Table~\ref{tab:parameters_mc3}.}
	\label{tab:three_cases}
\end{table}

\begin{figure}[h]
	\captionsetup[sub]{font=small}
	\centering
	\begin{subfigure}[b]{0.475\textwidth}
		\begin{tikzpicture}
			\node at (0,0) {	
%
%
\begin{tikzpicture}

\begin{axis}[%
width=6cm,
height=4cm,
at={(0cm,0cm)},
scale only axis,
unbounded coords=jump,
xmin=0,
xmax=500,
xlabel style={font=\color{white!15!black}},
xlabel={Time [d]},
ymin=160,
ymax=270,
ylabel style={font=\color{white!15!black}},
ylabel={Mass [g]},
axis background/.style={fill=white},
legend style={legend cell align=left, align=left, draw=white!15!black}
]
\addplot [color=blue, line width=2.0pt]
  table[row sep=crcr]{%
0	248.94\\
0.5	247\\
0.638318	246.78\\
0.868834	247.79\\
1.03896	249.15\\
1.08886	249.51\\
1.17493	250.02\\
1.42765	251\\
1.73721	251.69\\
2.19611	252.06\\
2.82297	251.88\\
3.71794	250.83\\
4.99308	248.46\\
6.84533	244.13\\
9.57332	236.97\\
13.679	226.04\\
20.0215	211.27\\
29.9597	195.67\\
33.3593	191.4\\
36.7589	188.05\\
40.1584	185.43\\
41.0083	184.82\\
41.8582	184.26\\
42.7081	183.73\\
43.558	183.24\\
44.4079	182.78\\
45.2578	182.35\\
53.7568	179.51\\
71.0769	176.42\\
102.623	174.02\\
152.623	172.66\\
202.623	172.13\\
252.623	171.92\\
302.623	171.82\\
352.623	171.78\\
402.623	171.76\\
452.623	171.75\\
502.623	171.74\\
552.623	171.74\\
602.623	171.74\\
652.623	171.74\\
702.623	171.74\\
752.623	171.74\\
802.623	171.74\\
};

\addplot [color=blue, line width=2.0pt, draw=none, mark size=1.0pt, mark=triangle, mark options={solid, blue}]
  table[row sep=crcr]{%
0	248.94\\
0.5	247\\
0.638318	246.78\\
0.868834	247.79\\
1.03896	249.15\\
1.08886	249.51\\
1.17493	250.02\\
1.42765	251\\
1.73721	251.69\\
2.19611	252.06\\
2.82297	251.88\\
3.71794	250.83\\
4.99308	248.46\\
6.84533	244.13\\
9.57332	236.97\\
13.679	226.04\\
20.0215	211.27\\
29.9597	195.67\\
33.3593	191.4\\
36.7589	188.05\\
40.1584	185.43\\
41.0083	184.82\\
41.8582	184.26\\
42.7081	183.73\\
43.558	183.24\\
44.4079	182.78\\
45.2578	182.35\\
53.7568	179.51\\
71.0769	176.42\\
102.623	174.02\\
152.623	172.66\\
202.623	172.13\\
252.623	171.92\\
302.623	171.82\\
352.623	171.78\\
402.623	171.76\\
452.623	171.75\\
502.623	171.74\\
552.623	171.74\\
602.623	171.74\\
652.623	171.74\\
702.623	171.74\\
752.623	171.74\\
802.623	171.74\\
};


\end{axis}

\end{tikzpicture}%
 };
			\node at(0.5,1.0){ \includegraphics[width=4.5cm]{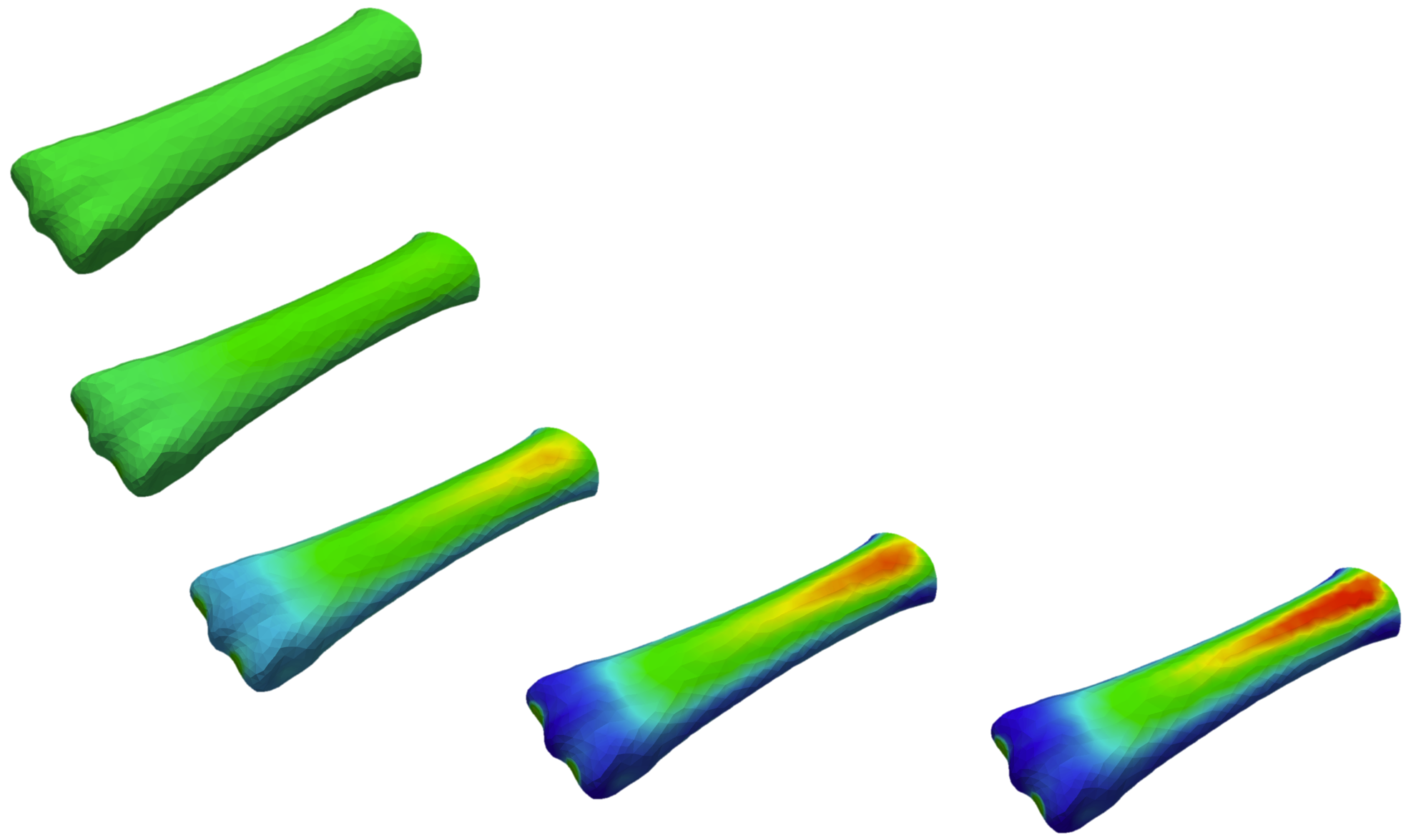}};
			\end{tikzpicture}
		\caption{ Case 1. }
		\label{fig:mc3_density}
	\end{subfigure}
	\hfill
	\begin{subfigure}[b]{0.475\textwidth}  
		\begin{tikzpicture}
			\node at (0,0) {	
%
%
\begin{tikzpicture}

\begin{axis}[%
width=6cm,
height=4cm,
at={(1.011in,0.723in)},
scale only axis,
unbounded coords=jump,
xmin=0,
xmax=100,
xlabel style={font=\color{white!15!black}},
xlabel={Time [d]},
ymin=160,
ymax=270,
ylabel style={font=\color{white!15!black}},
ylabel={Mass [g]},
axis background/.style={fill=white},
legend style={legend cell align=left, align=left, draw=white!15!black}
]
\addplot [color=blue, line width=2.0pt]
  table[row sep=crcr]{%
0	248.94\\
0.5	247\\
0.638318	246.78\\
0.868834	247.79\\
1.03896	249.15\\
1.08886	249.51\\
1.17493	250.02\\
1.42765	251\\
1.73721	251.69\\
2.19611	252.06\\
2.82297	251.88\\
3.71794	250.83\\
4.99308	248.46\\
6.84533	244.13\\
9.57332	236.97\\
13.679	226.04\\
20.0215	211.27\\
22.5061	206.03\\
24.9906	201.4\\
27.4751	197.39\\
28.0963	196.44\\
28.7174	195.53\\
29.3385	194.67\\
29.9597	193.86\\
30.5808	193.1\\
31.2019	192.38\\
32.7548	190.81\\
34.3076	189.44\\
35.8604	188.23\\
37.4133	187.16\\
38.9661	186.2\\
42.8034	184.31\\
46.6407	182.77\\
50.4781	181.48\\
54.3154	180.39\\
58.1527	179.47\\
60.1746	179.02\\
62.1966	178.61\\
62.7021	178.51\\
63.2075	178.41\\
63.713	178.31\\
64.2185	178.22\\
64.724	178.12\\
65.2295	178.03\\
65.735	177.94\\
66.2405	177.86\\
66.7459	177.77\\
67.2514	177.69\\
67.7569	177.61\\
72.8118	176.9\\
81.5878	175.97\\
90.3637	175.23\\
92.5577	175.06\\
};

\addplot [color=blue, line width=2.0pt, draw=none, mark size=1.0pt, mark=triangle, mark options={solid, blue}]
  table[row sep=crcr]{%
0	248.94\\
0.5	247\\
0.638318	246.78\\
0.868834	247.79\\
1.03896	249.15\\
1.08886	249.51\\
1.17493	250.02\\
1.42765	251\\
1.73721	251.69\\
2.19611	252.06\\
2.82297	251.88\\
3.71794	250.83\\
4.99308	248.46\\
6.84533	244.13\\
9.57332	236.97\\
13.679	226.04\\
20.0215	211.27\\
22.5061	206.03\\
24.9906	201.4\\
27.4751	197.39\\
28.0963	196.44\\
28.7174	195.53\\
29.3385	194.67\\
29.9597	193.86\\
30.5808	193.1\\
31.2019	192.38\\
32.7548	190.81\\
34.3076	189.44\\
35.8604	188.23\\
37.4133	187.16\\
38.9661	186.2\\
42.8034	184.31\\
46.6407	182.77\\
50.4781	181.48\\
54.3154	180.39\\
58.1527	179.47\\
60.1746	179.02\\
62.1966	178.61\\
62.7021	178.51\\
63.2075	178.41\\
63.713	178.31\\
64.2185	178.22\\
64.724	178.12\\
65.2295	178.03\\
65.735	177.94\\
66.2405	177.86\\
66.7459	177.77\\
67.2514	177.69\\
67.7569	177.61\\
72.8118	176.9\\
81.5878	175.97\\
90.3637	175.23\\
92.5577	175.06\\
};

\addplot [color=blue, line width=2.0pt, mark size=1.0pt, mark=triangle, mark options={solid, blue}]
  table[row sep=crcr]{%
nan	nan\\
};

\end{axis}

\end{tikzpicture}%
 };
			\node at(1.5,1.0){ \includegraphics[width=2cm]{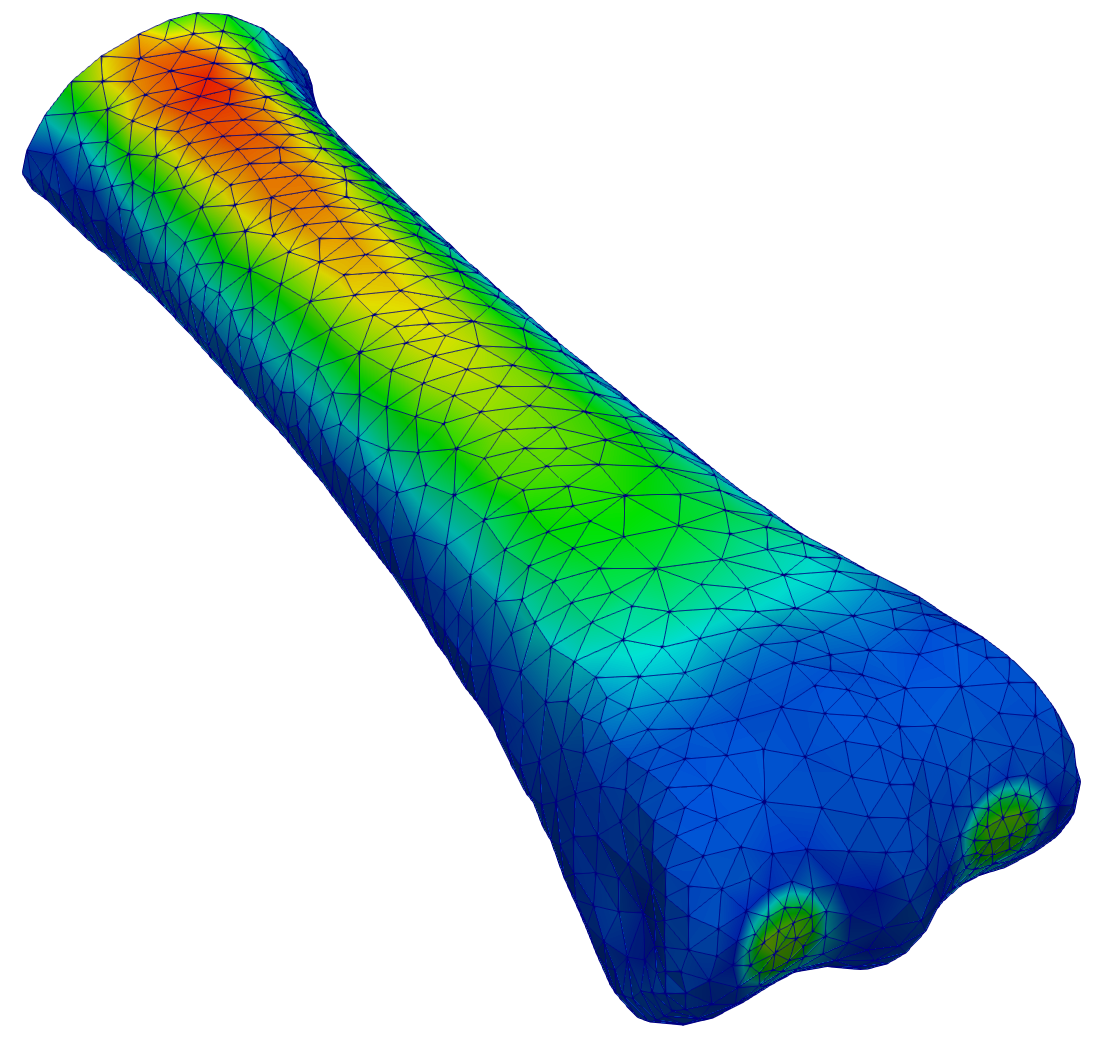}};
			\end{tikzpicture}

		\caption{ Case 2. }    
		\label{fig:density_bell1}
	\end{subfigure}
	\vskip\baselineskip
	\begin{subfigure}[b]{0.475\textwidth}   
		
		\begin{tikzpicture}
			\node at (0,0) {	
%
%
\begin{tikzpicture}

\begin{axis}[%
  width=6cm,
  height=4cm,
at={(1.011in,0.723in)},
scale only axis,
unbounded coords=jump,
xmin=0,
xmax=1700,
xlabel style={font=\color{white!15!black}},
xlabel={Time [d]},
ymin=240,
ymax=270,
ylabel style={font=\color{white!15!black}},
ylabel={Mass [g]},
axis background/.style={fill=white},
legend style={legend cell align=left, align=left, draw=white!15!black}
]
\addplot [color=blue, line width=2.0pt]
  table[row sep=crcr]{%
0	248.94\\
0.5	248.36\\
0.625	248.4\\
0.75	248.75\\
0.875	249.53\\
1	250.79\\
1.0803	251.51\\
1.15271	252.1\\
1.41315	253.69\\
1.70451	255.06\\
2.1455	256.56\\
2.73899	257.99\\
3.58705	259.36\\
4.78935	260.58\\
6.53053	261.57\\
9.08236	262.23\\
12.8971	262.46\\
18.6862	262.19\\
27.8059	261.39\\
43.2273	260.08\\
71.3203	258.31\\
121.32	256.23\\
171.32	254.7\\
221.32	253.48\\
271.32	252.48\\
321.32	251.62\\
371.32	250.88\\
421.32	250.22\\
471.32	249.63\\
521.32	249.1\\
571.32	248.61\\
621.32	248.16\\
671.32	247.74\\
721.32	247.35\\
771.32	246.98\\
821.32	246.64\\
871.32	246.31\\
921.32	246\\
971.32	245.7\\
1021.32	245.41\\
1071.32	245.14\\
1121.32	244.88\\
1171.32	244.63\\
1221.32	244.38\\
1271.32	244.15\\
1321.32	243.92\\
1371.32	243.7\\
1421.32	243.49\\
1471.32	243.28\\
1521.32	243.08\\
1571.32	242.89\\
1621.32	242.7\\
};

\addplot [color=blue, line width=2.0pt, draw=none, mark size=1.0pt, mark=triangle, mark options={solid, blue}]
  table[row sep=crcr]{%
0	248.94\\
0.5	248.36\\
0.625	248.4\\
0.75	248.75\\
0.875	249.53\\
1	250.79\\
1.0803	251.51\\
1.15271	252.1\\
1.41315	253.69\\
1.70451	255.06\\
2.1455	256.56\\
2.73899	257.99\\
3.58705	259.36\\
4.78935	260.58\\
6.53053	261.57\\
9.08236	262.23\\
12.8971	262.46\\
18.6862	262.19\\
27.8059	261.39\\
43.2273	260.08\\
71.3203	258.31\\
121.32	256.23\\
171.32	254.7\\
221.32	253.48\\
271.32	252.48\\
321.32	251.62\\
371.32	250.88\\
421.32	250.22\\
471.32	249.63\\
521.32	249.1\\
571.32	248.61\\
621.32	248.16\\
671.32	247.74\\
721.32	247.35\\
771.32	246.98\\
821.32	246.64\\
871.32	246.31\\
921.32	246\\
971.32	245.7\\
1021.32	245.41\\
1071.32	245.14\\
1121.32	244.88\\
1171.32	244.63\\
1221.32	244.38\\
1271.32	244.15\\
1321.32	243.92\\
1371.32	243.7\\
1421.32	243.49\\
1471.32	243.28\\
1521.32	243.08\\
1571.32	242.89\\
1621.32	242.7\\
};

\addplot [color=blue, line width=2.0pt, mark size=1.0pt, mark=triangle, mark options={solid, blue}]
  table[row sep=crcr]{%
nan	nan\\
};

\end{axis}
\end{tikzpicture}%
 };
			\node at(1.5,1.0){ \includegraphics[width=2cm]{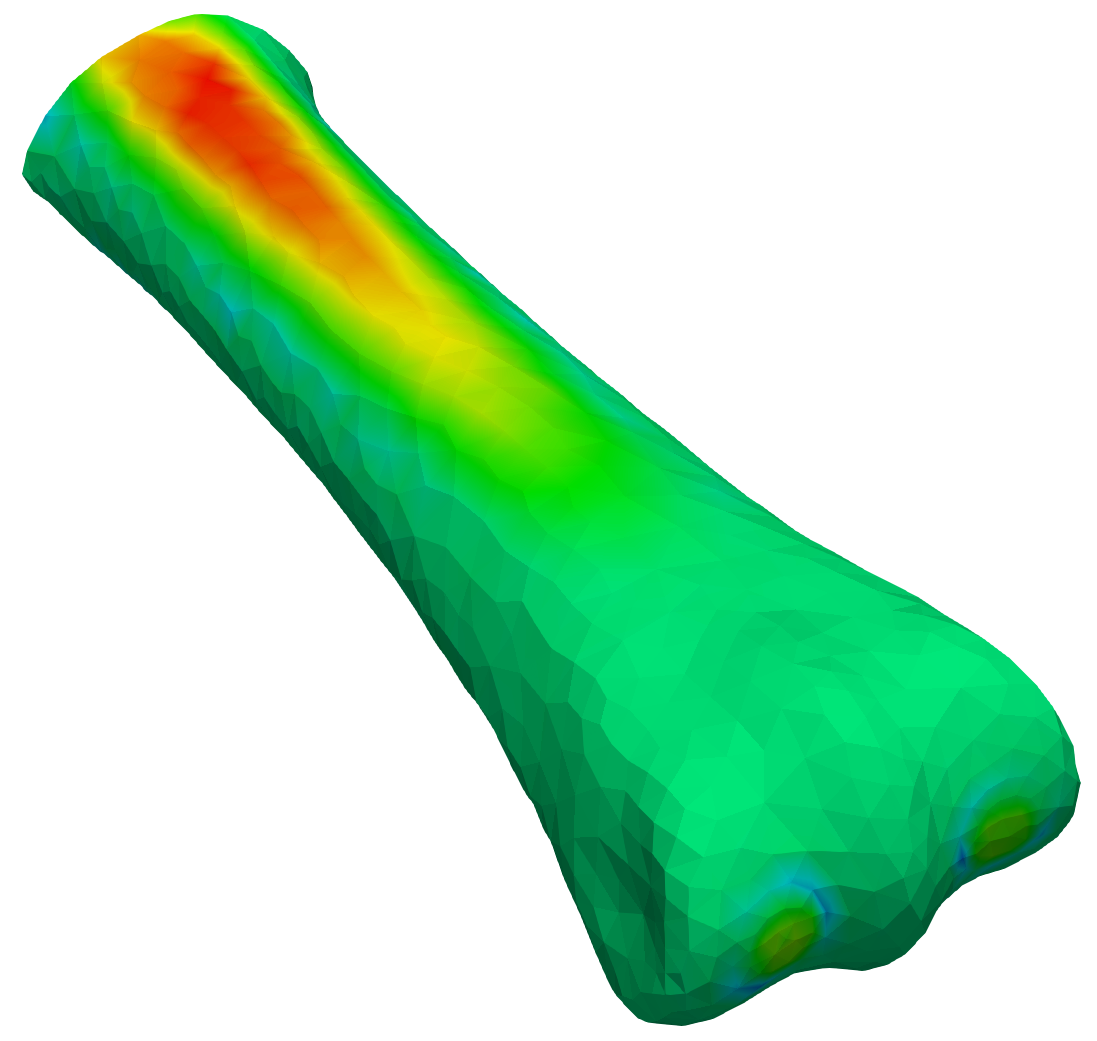}};
			\end{tikzpicture}
		\caption{ Case 3. }    
		\label{fig:density_bell2}
	\end{subfigure}
	\quad
	\begin{subfigure}[b]{0.475\textwidth}   
		\begin{tikzpicture}
			\node at(0,0){ } ;
			\node at(3,4.5){  \def\svgwidth{4cm} 	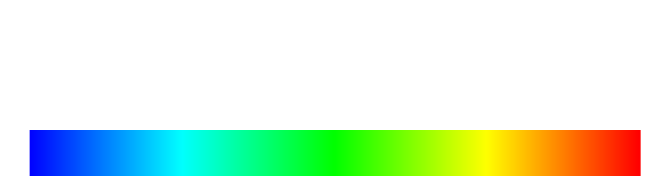  } ;
		\end{tikzpicture}
	\end{subfigure}
	\caption{Change in bone mass over time for 3 cases (see Table~\ref{tab:three_cases}). Density distribution contours in 3rd metacarpal bone at five snapshots in time in (a) and at the last converged step for (b) and (c).} 
	\label{fig:density_changes_}
\end{figure}

Results of Case~1 are presented in Figure~\ref{fig:mc3_density} where density maps at five different points in time $(\text{0, 10, 40, 100, 700})$~$[\text d]$ are visualised. 
Significant densification occurred immediately after reaching the maximum level of the loading, particularly in the proximity of the applied forces, associated with high levels of strain energy. Conversely, areas with low levels of strain energy experience a reduction in density.
After 100~$[{\text d}]$, biological equilibrium was achieved and no further changes in density took place. 
The resulting maximum density is 2.8~$[{\text {g/cm}}^3]$ and the minimum is close to zero.

The maximum density for Case~1 is noticeably higher than in the actual equine bones~\citep{yamada2015experimental} and the minimum density is unrealistically close to zero. This justifies the proposed bell shape function (Eq.~(\ref{eq:bell_function})) used for the next two analyses (Case~2 and Case~3). 

The results for Case~2 are plotted in Figure~\ref{fig:density_bell1}.
The last converged step takes place at $t=93$~$[{\text{d}}]$. 
By setting a high value of $b$, the transition between densities is very sharp and the algorithm encounters convergence difficulties, even with adaptive time-stepping, and biological equilibrium cannot be achieved in this case. 
However, by observing the range of densities obtained, it is evident that they slowly converge to the same solution as Case~1 (Figure~\ref{fig:mc3_density}).

For Case~3, a more moderate value for the exponent in the bell function was chosen along with a narrower density range than those chosen for Case~2 (see Table~\ref{tab:three_cases}).
The plot presented in Figure~\ref{fig:density_bell2} demonstrates how these values influence the results of the analysis. 
It is evident that with a much lower value for exponent, $b$, the algorithm no longer has problems converging. 
Furthermore,  reducing the range between the upper and lower bounds of density has a significant impact on the results. 
The dense cortical shaft on the dorsal side of the bone is less dense and covers a much larger region. 
Furthermore, unrealistically low values of densities have been eliminated. 
However, as with the previous case, the overall solution converges to the same mass as in Case~1, albeit requiring significantly more time steps. \\

\subsection{Stress intensity calculations}
\label{sec:release_energy_rate}
To examine the calculation of configurational forces at the crack front in bodies with both homogeneous and heterogeneous density distributions, five numerical examples are presented.
First, a simple quasi-two-dimensional plate with homogeneous material distribution is considered.  The convergence study utilises an analytical solution as a reference.
Second, the proposed singularity elements are included for the same plate problem and their influence on the rate of convergence is presented. 
Third, the same problem is considered again but with a heterogeneous material distribution.
The final two examples demonstrate the calculation of configurational forces for a more representative bone.
 
\subsubsection{Finite plate with a horizontal crack}\label{sec:plate_section}
\begin{figure}[h!]
\begin{center}
\begin{tabular}{c}
{\def\svgwidth{15cm} 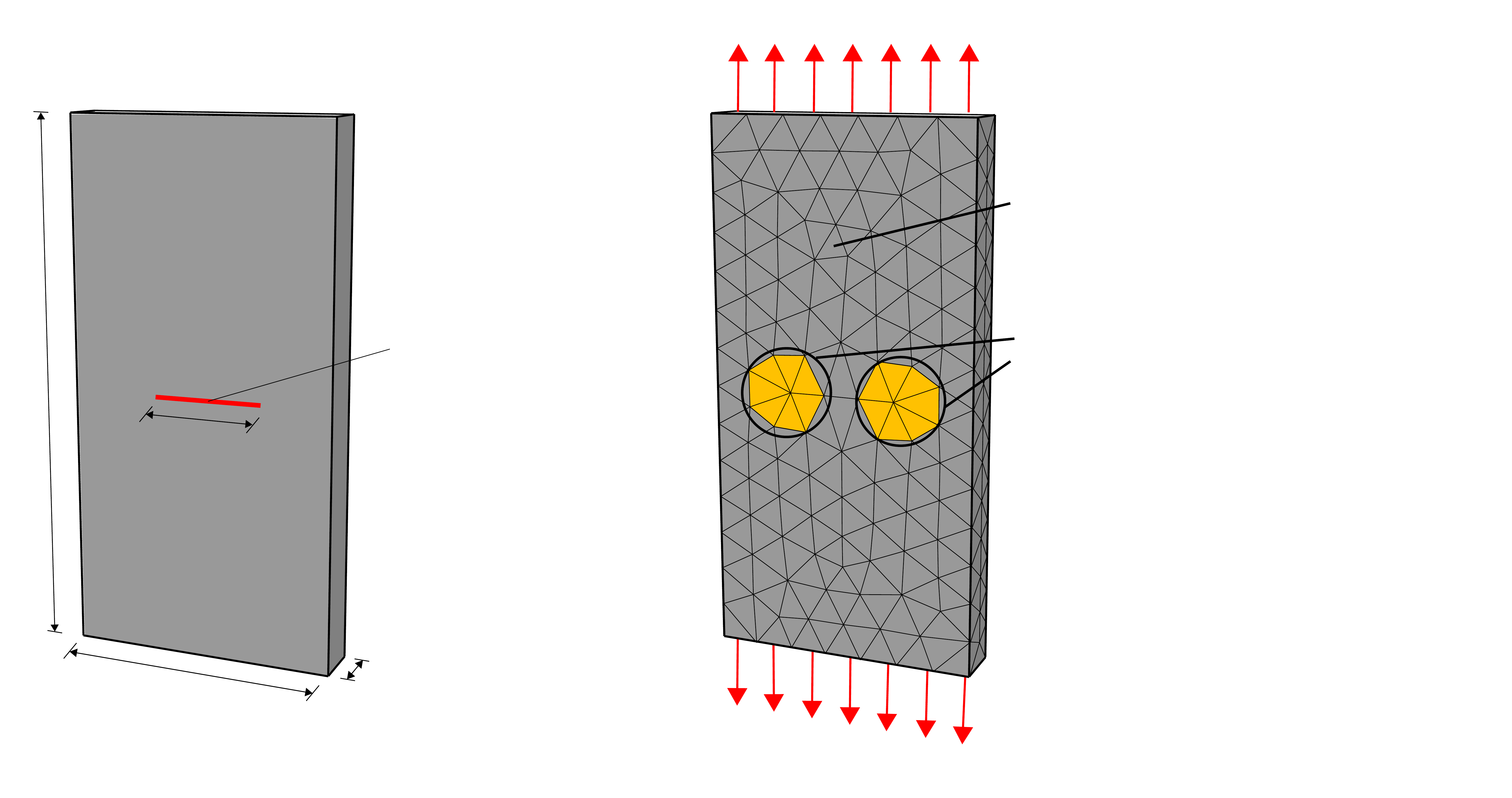}\\
\end{tabular}
\caption{Finite plate with a horizontal crack. a) Plate geometry with through thickness crack. b) Finite element mesh - grey elements have approximation order $p_{\rm g}$ and yellow elements have vertices at crack tip and have approximation order $p_{\rm l}+p_{\rm g}$. }
\label{fig:plate_load_mesh}
\end{center}
\end{figure}

A finite plate with height, $h_{\rm {pl}} = 10$, thickness $t_{\rm {pl}}=1$ and half width $b_{\rm{pl}} = 2.5$ and a horizontal through-thickness crack with half width $a_{\rm {pl}} = 1$, as presented in Figure~\ref{fig:plate_load_mesh}(a), is considered. All input parameters presented are dimensionless. 
The  plate is spatially discretised using 1384 tetrahedral elements and subjected to uniaxial stress in the longitudinal direction, as indicated in Figure~\ref{fig:plate_load_mesh}(b). 
Displacements are constrained on three vertices of the plate to prevent rigid body motion. 

The purpose of this analysis is to calculate the Mode I stress intensity factor $ K_{\rm I} $ directly from the configurational and compare with the analytical solution~\citep{rooke1976compendium} for an infinite plate:
\begin{equation}\label{eq:frac_analytical}
K_{\rm I}=\sigma \sqrt{\pi a_{\rm {pl}}} \left[  \frac{1 - \frac{a_{\rm {pl}}}{2b_{\rm {pl}}} + 0.326 (\frac{a_{\rm {pl}}}{b_{\rm {pl}}})^2 }{\sqrt{1-\frac{a_{\rm {pl}}}{b_{\rm {pl}}}}}  \right]
\end{equation}
where $\sigma $ is the applied stress. 
Young's modulus $E$ and Poisson's ratio $\nu$ are $1000$ and $0.3$, respectively.

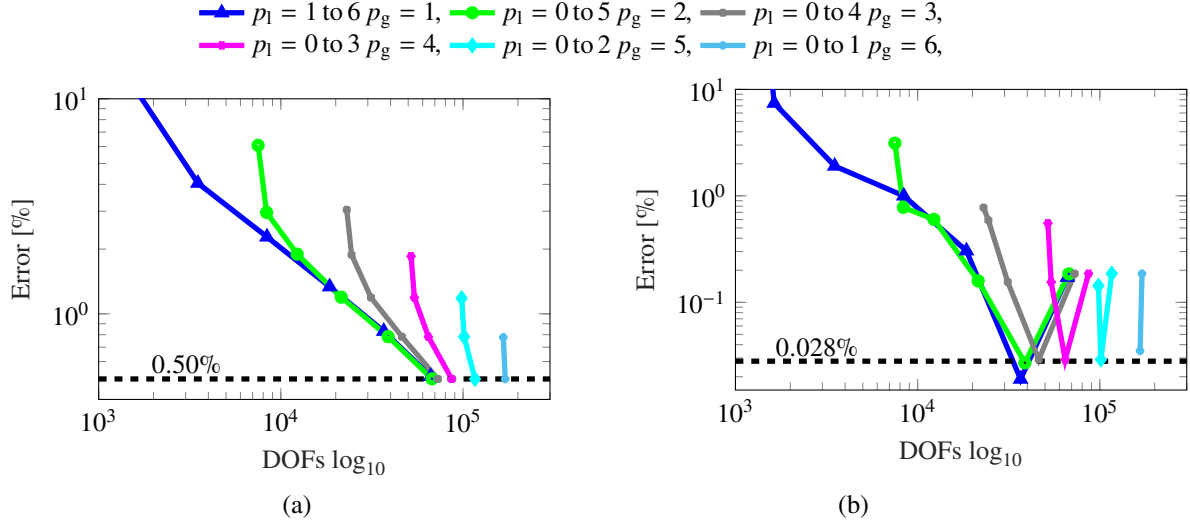
\begin{figure}[h]
	\centering
%
%
\definecolor{mycolor1}{rgb}{1.00000,0.00000,1.00000}%
\definecolor{mycolor2}{rgb}{0.00000,1.00000,1.00000}%
\definecolor{mycolor3}{rgb}{0.30100,0.74500,0.93300}%
\begin{tikzpicture}

\begin{axis}[%
width=10cm,
height=2cm,
hide axis,
axis background/.style={fill=white},
legend style={legend cell align=left, align=left, fill=none, draw=none},
legend columns=3,
ymin=-10,
ymax=-1,
xmin=-10,
xmax=-1,
legend image post style={sharp plot}
]

\addplot [color=blue, line width=2.0pt, mark size=1.5pt, mark=triangle, mark options={solid, blue}]
  table[row sep=crcr]{%
0 0\\
};
\addlegendentry{$p_{\rm l}= 1 \, \mathrm{to} \, 6 \: p_{\rm g}=1,$}

\addplot [color=green, line width=2.0pt, mark size=1.5pt, mark=o, mark options={solid, green}]
  table[row sep=crcr]{%
0 0\\
};
\addlegendentry{$p_{\rm l}= 0 \, \mathrm{to} \, 5 \: p_{\rm g}=2,$}

\addplot [color=gray, line width=2.0pt, mark size=1.5pt, mark=+, mark options={solid, gray}]
  table[row sep=crcr]{%
  0 0\\
};
\addlegendentry{$p_{\rm l}= 0 \, \mathrm{to} \, 4 \: p_{\rm g}=3,$}

\addplot [color=mycolor1, line width=2.0pt, mark size=1.5pt, mark=asterisk, mark options={solid, mycolor1}]
  table[row sep=crcr]{%
  0 0\\
};
\addlegendentry{$p_{\rm l}= 0 \, \mathrm{to} \, 3 \: p_{\rm g}=4,$}

\addplot [color=mycolor2, line width=2.0pt, mark size=1.5pt, mark=diamond, mark options={solid, mycolor2}]
  table[row sep=crcr]{%
  0 0\\
};
\addlegendentry{$p_{\rm l}= 0 \, \mathrm{to} \, 2 \: p_{\rm g}=5,$}

\addplot [color=mycolor3, line width=2.0pt, mark size=1.5pt, mark=x, mark options={solid, mycolor3}]
  table[row sep=crcr]{%
  0 0\\
};
\addlegendentry{$p_{\rm l}= 0 \, \mathrm{to} \, 1 \: p_{\rm g}=6,$}

\addplot [color=black, dashed, line width=2.0pt, forget plot]
  table[row sep=crcr]{%
  0 0\\
};

\end{axis}
\end{tikzpicture}%
	\begin{minipage}{.45\textwidth}
		\hspace{-0.5cm}
		\begin{tikzpicture}
			\node at (0,0)   {		
%
%
\definecolor{mycolor1}{rgb}{1.00000,0.00000,1.00000}%
\definecolor{mycolor2}{rgb}{0.00000,1.00000,1.00000}%
\definecolor{mycolor3}{rgb}{0.30100,0.74500,0.93300}%
\begin{tikzpicture}

\begin{axis}[%
width=6cm,
height=4cm,
scale only axis,
xmode=log,
xmin=1000,
xmax=300000,
xminorticks=true,
xlabel style={font=\color{white!15!black}},
xlabel={DOFs $\mathrm{log}_{10}$},
ymode=log,
ymin=0.4,
ymax=10,
yminorticks=true,
ylabel style={font=\color{white!15!black}},
ylabel={Error [\%]},
axis background/.style={fill=white},
legend style={legend cell align=left, align=left, fill=none, draw=none},
legend columns=2
]
\addplot [color=blue, line width=2.0pt, mark size=1.5pt, mark=triangle, mark options={solid, blue}]
  table[row sep=crcr]{%
1219	22.692\\
1633	10.729\\
3505	4.053\\
8392	2.279\\
18529	1.337\\
36721	0.83\\
66427	0.512\\
};

\addplot [color=green, line width=2.0pt, mark size=1.5pt, mark=o, mark options={solid, green}]
  table[row sep=crcr]{%
7501	6.075\\
8353	2.96\\
12286	1.895\\
21475	1.195\\
38785	0.784\\
67582	0.499\\
};

\addplot [color=gray, line width=2.0pt, mark size=1.5pt, mark=+, mark options={solid, gray}]
  table[row sep=crcr]{%
22999	3.048\\
24445	1.883\\
31237	1.191\\
46189	0.783\\
72775	0.498\\
};

\addplot [color=mycolor1, line width=2.0pt, mark size=1.5pt, mark=asterisk, mark options={solid, mycolor1}]
  table[row sep=crcr]{%
51865	1.854\\
54061	1.191\\
64510	0.783\\
86686	0.498\\
};

\addplot [color=mycolor2, line width=2.0pt, mark size=1.5pt, mark=diamond, mark options={solid, mycolor2}]
  table[row sep=crcr]{%
98251	1.181\\
101353	0.783\\
116257	0.498\\
};

\addplot [color=mycolor3, line width=2.0pt, mark size=1.5pt, mark=x, mark options={solid, mycolor3}]
  table[row sep=crcr]{%
166309	0.778\\
170473	0.498\\
};

\addplot [color=black, dashed, line width=2.0pt, forget plot]
  table[row sep=crcr]{%
1000	0.498\\
300000	0.498\\
};
\end{axis}
\end{tikzpicture}%
 };
			\node at (-1.2,-1.1){$0.50\%$};
			\node at (0,0) {};
			\end{tikzpicture}
	\end{minipage}%
	\quad \quad \quad
	\begin{minipage}{.45\textwidth}
		\vspace{-0.25cm}
		\begin{tikzpicture}
			\node at (0,0){	
%
%
\definecolor{mycolor1}{rgb}{1.00000,0.00000,1.00000}%
\definecolor{mycolor2}{rgb}{0.00000,1.00000,1.00000}%
\definecolor{mycolor3}{rgb}{0.30100,0.74500,0.93300}%
\begin{tikzpicture}

\begin{axis}[%
width=6cm,
height=4cm,
scale only axis,
xmode=log,
xmin=1000,
xmax=300000,
xminorticks=true,
xlabel style={font=\color{white!15!black}},
xlabel={DOFs $\mathrm{log}_{10}$},
ymode=log,
ymin=0.015,
ymax=10,
yminorticks=true,
ylabel style={font=\color{white!15!black}},
ylabel={Error [\%]},
axis background/.style={fill=white},
legend style={legend cell align=left, align=left, fill=none, draw=none},
legend columns=2
]
\addplot [color=blue, line width=2.0pt, mark size=1.5pt, mark=triangle, mark options={solid, blue}]
  table[row sep=crcr]{%
1219	978.472\\
1633	7.424\\
3505	1.905\\
8392	0.999\\
18529	0.305\\
36721	0.019\\
66427	0.172\\
};

\addplot [color=green, line width=2.0pt, mark size=1.5pt, mark=o, mark options={solid, green}]
  table[row sep=crcr]{%
7501	3.128\\
8353	0.779\\
12286	0.601\\
21475	0.159\\
38785	0.027\\
67582	0.185\\
};

\addplot [color=gray, line width=2.0pt, mark size=1.5pt, mark=+, mark options={solid, gray}]
  table[row sep=crcr]{%
22999	0.776\\
24445	0.589\\
31237	0.155\\
46189	0.029\\
72775	0.186\\
};

\addplot [color=mycolor1, line width=2.0pt, mark size=1.5pt, mark=asterisk, mark options={solid, mycolor1}]
  table[row sep=crcr]{%
51865	0.553\\
54061	0.155\\
64510	0.029\\
86686	0.186\\
};

\addplot [color=mycolor2, line width=2.0pt, mark size=1.5pt, mark=diamond, mark options={solid, mycolor2}]
  table[row sep=crcr]{%
98251	0.143\\
101353	0.029\\
116257	0.186\\
};

\addplot [color=mycolor3, line width=2.0pt, mark size=1.5pt, mark=x, mark options={solid, mycolor3}]
  table[row sep=crcr]{%
166309	0.035\\
170473	0.186\\
};

\addplot [color=black, dashed, line width=2.0pt, forget plot]
  table[row sep=crcr]{%
1000	0.028\\
300000	0.028\\
};
\end{axis}
\end{tikzpicture}%
 };
			\node at (-1.2,-1.0){$0.028\%$};
	\end{tikzpicture}
	\end{minipage}
				\begin{centering} (a)\hspace{7cm}(b) \end{centering}
	\caption{Convergence plot for stress intensity factor $K_{\rm I}$. Relative error (\%) versus no. of DOF (log10) for (a) using hierarchical approximation functions and (b) using singularity elements.}
			\label{fig:plate_conv}
\end{figure}

Hierarchical approximation functions allow for global and local p-refinment without changing the mesh.
In general, all tetrahedrons of the mesh have a global order of approximation, $p_{\rm g}$, with some elements subjected to local refinement of order $p_{\rm{l}}$.
All analyses presented were run using the same mesh with p-refinement varying from 1\textsuperscript{st}-order to 6\textsuperscript{th}-order so that $p_{\rm l} + p_{\rm g} \leq 7$.
The Mode I stress intensity factor, $K_{\rm I}$, was calculated directly from the output configurational forces as: 
\begin{equation}
K_{\rm I} = \sqrt{GE}
\end{equation}
where $G$ is the change of elastic strain energy per unit area of crack growth.
From Figure~\ref{fig:plate_conv}(a), it is evident that, for the same coarse mesh and number of nodes, the solution can improve drastically when the order of approximation is increased. 
The well known shear locking associated with first-order approximation is observed. 
The minimum error achieved is $0.50\%$ for all the cases with total order of approximation  $p_{\rm l} + p_{\rm g} = 7$. 
Therefore, it can be observed that using a low order of global approximation plus local p-refinement 
can achieve the same level of accuracy as using high order approximation globally, but with fewer degrees of freedom and lower computational cost.

Based on the results in Figure~\ref{fig:plate_conv}(b), it is evident that using singularity elements improves the convergence rate significantly and lowers the error by an order of magnitude, from 0.50\% down to 0.028\%. 
However, it can also be seen that for each combination of p-refinement, the error increases with further refinement after it reaches the minimum value. 
This suggests that the solution cannot be further improved by enhancing the order of approximation alone. 
Reducing the the elements size and increasing the plate width (to better replicate the infinite plate used to determine the analytical solution), the error would probably decrease further. 
Nevertheless, the results are considered sufficiently accurate for the purpose at hand. 

Overall, these results indicate that it is of great benefit to use the singularity elements, since they improve the accuracy of the solution with no extra cost.
Furthermore, the difference in execution time for the analysis with and without their inclusion was negligible. 

\subsubsection{Heterogeneous material}
So far the numerical examples have assumed homogenous material properties. Here we consider the effect of a heterogeneous density distribution. 
Considering the same problem of the finite plate with horizontal crack, a density field $\mathbf{\rho}(x,y,z) = 0.125y + 1$ is directly assigned to the integration (Gauss) points of each tetrahedral element.
As expected, configurational forces are induced at the crack tip under load and, as explained in Section \ref{sec:fem_fracture_prop}, these forces are influenced by the non-uniform density distribution. 
However, the stress intensity factor loses its meaning in the case of heterogeneous materials and there is no agreed approach to validate either configurational forces or stress intensity factors for such cases (except for the special case of functionally graded materials~\citep{kim2002finite}).

\begin{figure}[h]
	\centering
	\begin{tikzpicture}
		\node at (0,0){
%
%
\definecolor{mycolor1}{rgb}{1.00000,0.00000,1.00000}%
\begin{tikzpicture}

\begin{axis}[%
width=12cm,
height=8cm,
at={(1.011in,0.799in)},
scale only axis,
xmode=log,
xmin=1000,
xmax=500000,
xminorticks=true,
xlabel style={font=\color{white!15!black}},
xlabel={DOFs $\mathrm{log}_{10}$},
ymode=log,
ymin=0.2,
ymax=100,
yminorticks=true,
ylabel style={font=\color{white!15!black}},
ylabel={Error [\%]},
axis background/.style={fill=white},
legend style={legend cell align=left, align=left, fill=none, draw=none}
]
\addplot [color=blue, line width=2.0pt, mark size=2.0pt, mark=triangle, mark options={solid, blue}]
  table[row sep=crcr]{%
1693	45.01\\
8614	0.8301\\
25186	0.8503\\
55690	0.7077\\
104407	0.4907\\
175618	0.3138\\
};
\addlegendentry{$p_{\rm g}= 1 \, \mathrm{to} \, 6 \: \Delta a_{\rm pl}=0.00125$}

\addplot [color=green, line width=2.0pt, mark size=2.0pt, mark=o, mark options={solid, green}]
  table[row sep=crcr]{%
1663	422.4\\
8521	2.829\\
24970	0.8699\\
55270	0.8441\\
103681	0.6077\\
174463	0.431\\
};
\addlegendentry{$p_{\rm g}= 1 \, \mathrm{to} \, 6 \: \Delta a_{\rm pl}=0.0025$}

\addplot [color=gray, line width=2.0pt, mark size=2.0pt, mark=+, mark options={solid, gray}]
  table[row sep=crcr]{%
1645	48.14\\
8461	7.211\\
24817	1.686\\
54949	0.7857\\
103093	0.5297\\
173485	0.3529\\
};
\addlegendentry{$p_{\rm g}= 1 \, \mathrm{to} \, 6 \: \Delta a_{\rm pl}=0.005$}

\addplot [color=mycolor1, line width=2.0pt, mark size=2.0pt, mark=asterisk, mark options={solid, mycolor1}]
  table[row sep=crcr]{%
1672	29.09\\
8515	16.284\\
24898	0.962\\
55051	0.7467\\
103204	0.5492\\
173587	0.3919\\
};
\addlegendentry{$p_{\rm g}= 1 \, \mathrm{to} \, 6 \: \Delta a_{\rm pl}=0.01$}

\addplot [color=black, dashed, line width=2.0pt, forget plot]
  table[row sep=crcr]{%
1000	0.3\\
5000000	0.3\\
};
\end{axis}
\end{tikzpicture}%
};
		\node at (5,-0.0) {	\includegraphics[width=2cm]{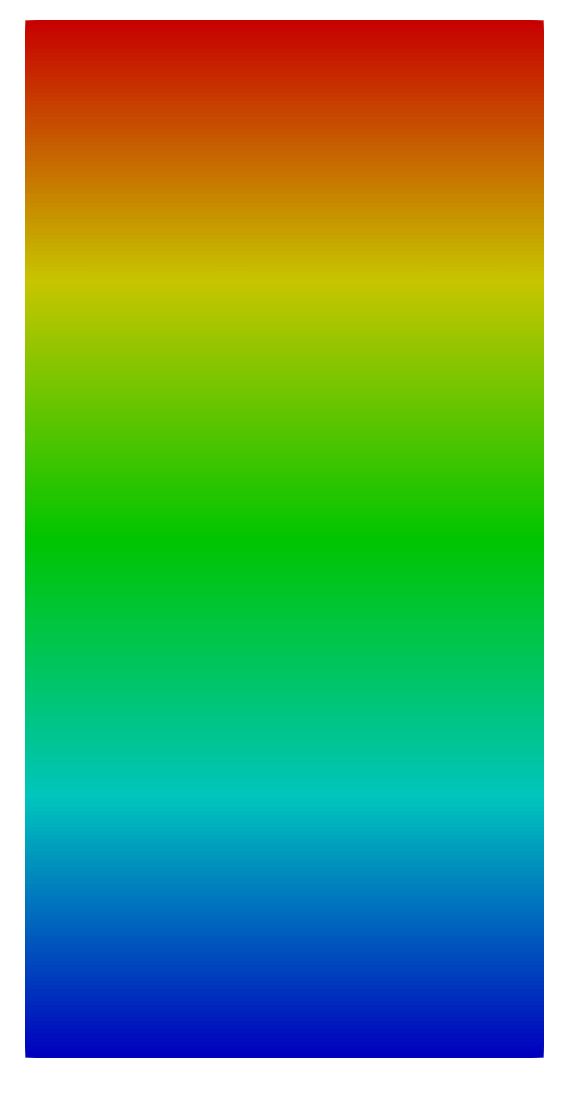}};
		\node at (-4.5,-2.8){$0.3\%$};
		\end{tikzpicture}
	\caption{Convergence plot for stress intensity factor $K_{\rm I}$ for heterogeneous density distribution. Relative error (\%) versus no. of DOF (log10).}
	\label{fig:covergencefdm}
\end{figure}
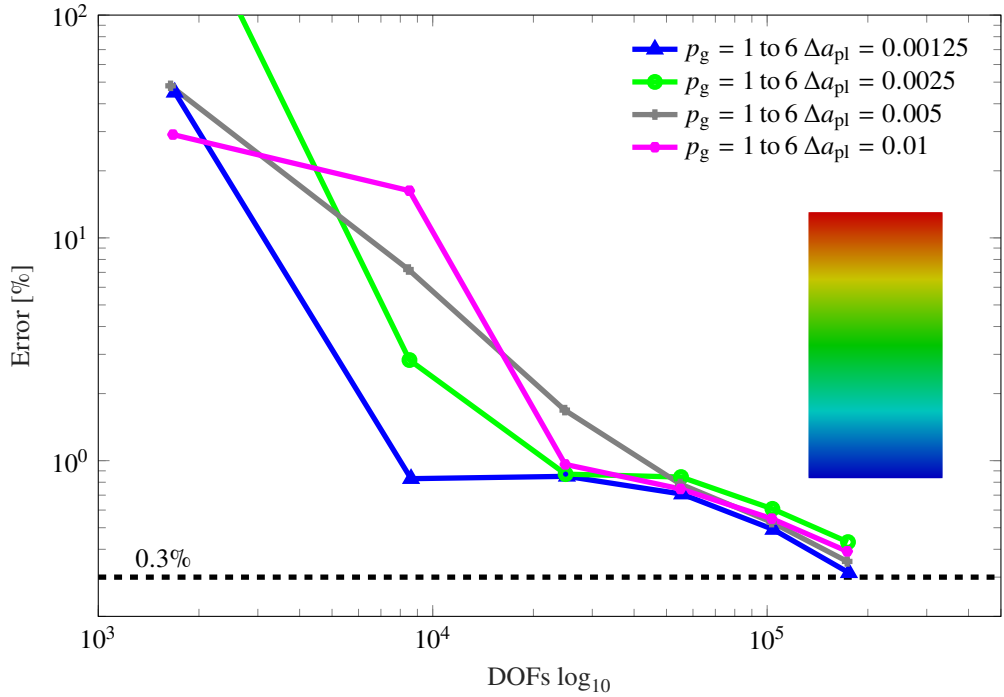

A straightforward verification can be performed by using a central difference numerical integration. The energy release rate for crack growth can be calculated as the change in elastic strain energy per unit area of crack growth~\citep{Griffith163}:
\begin{equation}
G = \frac{\partial \psi}{\partial a_{\rm {pl}}}
\end{equation}
where $\psi$ is the elastic energy of the system, and $a_{\rm{pl}}$ is the crack length. This derivative can be approximated as:
\begin{equation}
 \frac{\partial \psi}{\partial a_{\rm {pl}}} = \lim_{\Delta a_{\rm {pl}} \to 0} \frac{\psi(a_{\rm {pl}} + \Delta a_{\rm {pl}}) - \psi(a_{\rm {pl}} -\Delta a_{\rm {pl}})}{2\Delta a_{\rm {pl}}}
\end{equation}
where the elastic strain energies $\psi(a_{\rm {pl}} \pm \Delta a_{\rm {pl}})$ is obtained from two additional analyses with horizontal cracks of lengths: ($a + \Delta a_{\rm {pl}}$) and ($a - \Delta a_{\rm {pl}}$), where $\Delta a_{\rm {pl}}$ is a very small value. 
Next, knowing the resulting release energy with the crack length of $a_{\rm {pl}}$, a relative error can be calculated. 
Twenty-four analyses, for different levels of $p$~-~refinement and values of $\Delta a_{\rm {pl}}$, have been undertaken in order to determine the error in the release energy. 
The results are presented in Figure~\ref{fig:covergencefdm}, where it is apparent that the error in fracture energy release rate is converging to 0.3\% 
with increasing levels of refinement. It is worth noting that a similar level of accuracy was attained for the homogeneous case.  Achieving higher precision with this means of validation is difficult due to the accumulation of truncation, approximation and discretisation errors.
Therefore, it can be concluded that the proposed estimation of fracture energy release rate for heterogeneous materials is obtained with a satisfactory level of accuracy.

\subsubsection{Fracture energy release rate for metacarpal bone} \label{sec:mc3_release_eng}
\begin{figure}[h]
	\centering
		\def\svgwidth{10cm}
		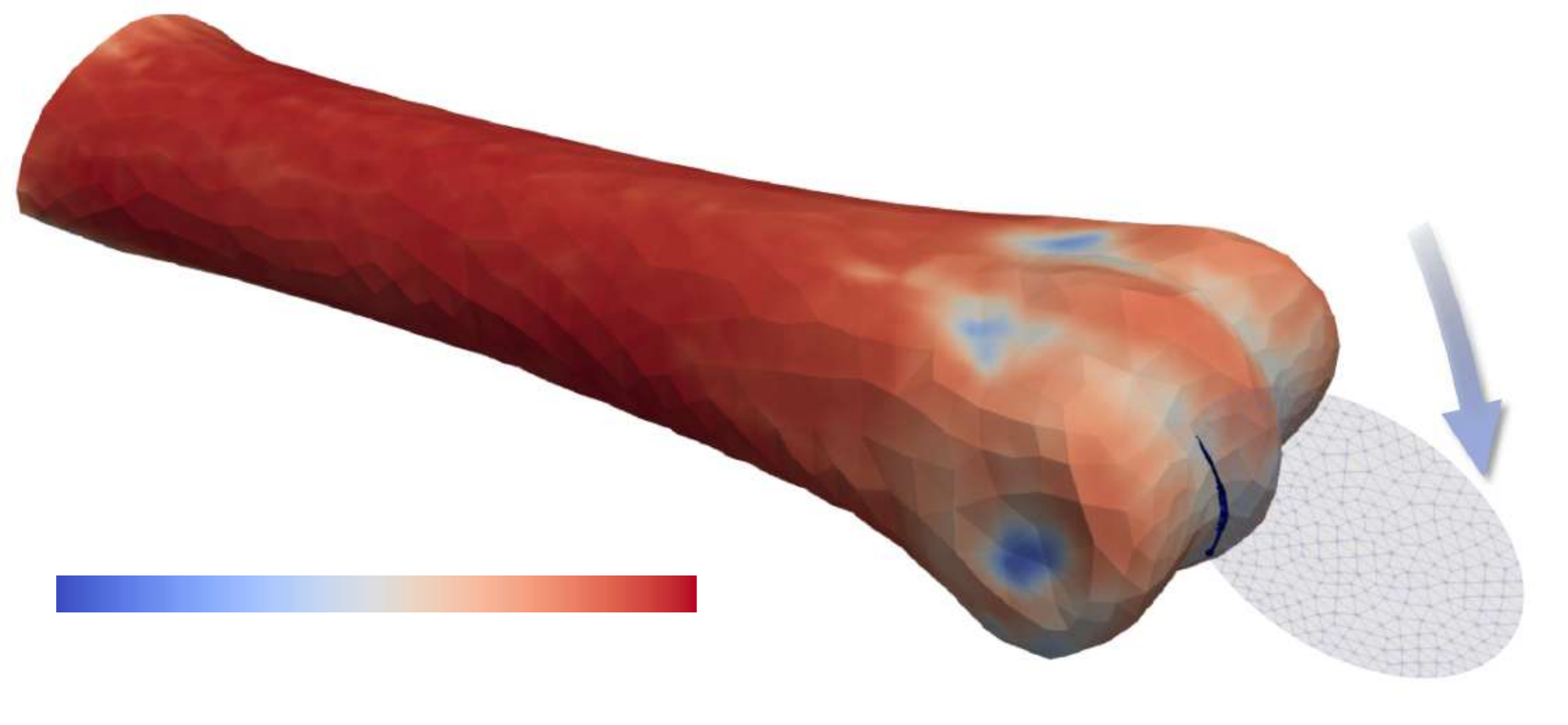
	\caption{Bone geometry with density mapped from CT using MWLS. Initial crack introduced by cutting the mesh with a circular surface.}
	\label{fig:bone_ct_mesh_cut}
\end{figure}

\begin{figure}[h!]
	\centering
	\def\svgwidth{12cm}
	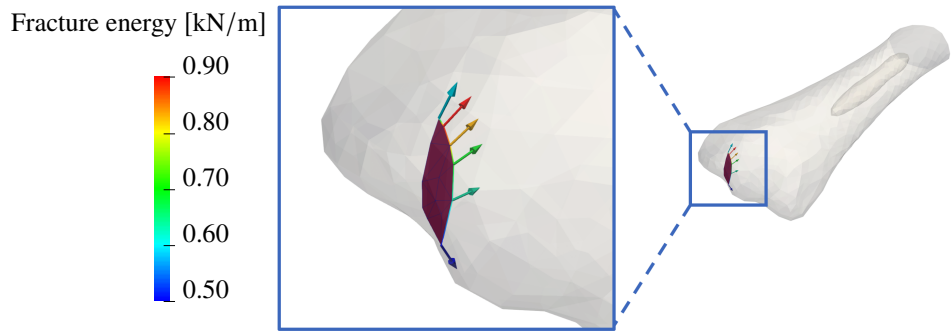
	\caption{Crack surface and configurational forces at the crack front.}
	\label{fig:crackfrontforce}
\end{figure}

\begin{figure}[h!]
	\centering
%
%
\definecolor{mycolor1}{rgb}{1.00000,0.00000,1.00000}%
\begin{tikzpicture}

\begin{axis}[%
width=10cm,
height=6cm,
at={(1.011in,0.799in)},
scale only axis,
xmode=log,
xmin=3000,
xmax=600000,
xminorticks=true,
xlabel style={font=\color{white!15!black}},
xlabel={DOFs $\mathrm{log}_{10}$},
ymin=0.5,
ymax=1,
ylabel style={font=\color{white!15!black}},
ylabel={Fracture energy release rate [$\mathrm{kN} / \mathrm{m}$]},
axis background/.style={fill=white},
legend style={at={(0.97,0.03)}, anchor=south east, legend cell align=left, align=left, fill=none, draw=none}
]
\addplot [color=blue, line width=2.0pt, mark size=2.0pt, mark=triangle, mark options={solid, blue}]
  table[row sep=crcr]{%
4957	0.596\\
6946	0.783\\
12787	0.865\\
35248	0.883\\
46615	0.893\\
103867	0.895\\
122743	0.9\\
};
\addlegendentry{$p_{\rm l}= 1 \, \mathrm{to} \, 3, \: p_{\rm g}=1 \, \mathrm{to} \, 3,$ coarse mesh}

\addplot [color=green, line width=2.0pt, mark size=2.0pt, mark=o, mark options={solid, green}]
  table[row sep=crcr]{%
7816	0.701\\
10405	0.809\\
17692	0.876\\
55684	0.9\\
69757	0.898\\
166900	0.906\\
190165	0.905\\
};
\addlegendentry{$p_{\rm l}= 2 \, \mathrm{to} \, 3, \: p_{\rm g}=1 \, \mathrm{to} \, 3,$ base mesh}

\addplot [color=mycolor1, line width=2.0pt, mark size=2.0pt, mark=+, mark options={solid, mycolor1}]
  table[row sep=crcr]{%
15313	0.69\\
18406	0.821\\
27412	0.861\\
108895	0.888\\
126496	0.896\\
336172	0.898\\
365473	0.899\\
};
\addlegendentry{$p_{\rm l}= 2 \, \mathrm{to} \, 3, \: p_{\rm g}=1 \, \mathrm{to} \, 3,$ fine mesh}

\addplot [color=black, dashed, line width=2.0pt, forget plot]
  table[row sep=crcr]{%
3000	0.9\\
600000	0.9\\
};
\end{axis}

\end{tikzpicture}%
	\caption{Convergence plot of fracture energy release rate versus no of DOF (log10) for subsequent discretisations and $p$~-~refinements.}
	\label{fig:max_g1_convergece}
\end{figure}
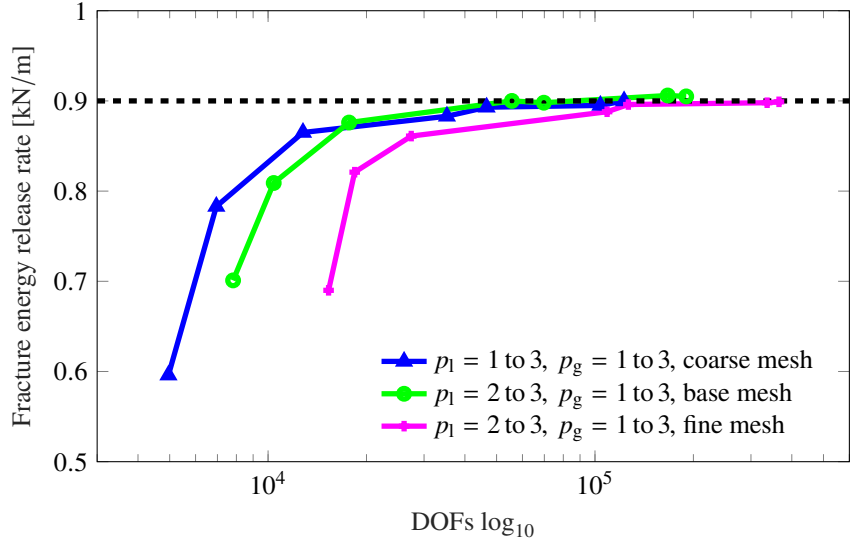

This numerical example considers the same bone as presented in Section~\ref{sec:numerical_examples:bone_adap}. 
An initial crack was generated in the mesh using a cutting plane, as shown in Figure~\ref{fig:bone_ct_mesh_cut}. A notch is situated at the origin of the most common location of a lateral condyle fracture~\citep{jacklin2012frequency}. 
The numerical analyses were undertaken using three meshes consisted of 6069, 10032 and 21189 tetrahedrons and repeated for 1\textsuperscript{st}, 2\textsuperscript{nd} and 3\textsuperscript{rd}-order of global $p$~-~refinement and local $p$~-~refinement at the crack tip. 
Boundary conditions and material parameters remain the same as in Table~\ref{tab:parameters_mc3}. 
Using a $\mathrm {K_2 HPO_4}$ calibration phantom, grey scale values from CT scans are converted to bone mineral density using five tubes with reference densities. 
The mechanical material properties were mapped onto the integration points of the mesh of the metacarpal bone using the MWLS method described earlier. 
The application of load induces configurational forces at the crack front, as shown in Figure~\ref{fig:crackfrontforce}. 
The direction of the vectors also indicates the direction of crack propagation.
The values of numerically predicted maximal nodal fracture energy release rates in Mode I (crack opening) for subsequent meshes are plotted in Figure~\ref{fig:max_g1_convergece}. 
It can be seen that, for the same mesh, as the order of approximation increases, the energy release rate converges. 

A crack will propagate when the energy release rate $G$ equals the material's resistance to crack extension, $g_{\rm c}$. Assuming $g_{\rm c} = 2.0\,[\mathrm{ kJ/m^2}]$~\citep{gasser2007numerical} it can be estimated that this particular metacarpal bone with this initial crack can sustain loading of approximately 2.2 times greater before a fracture starts to propagate. 

\subsubsection{Fracture energy release rate for adapted bone}
\begin{figure}[h]
	\centering
			\def\svgwidth{9cm}
		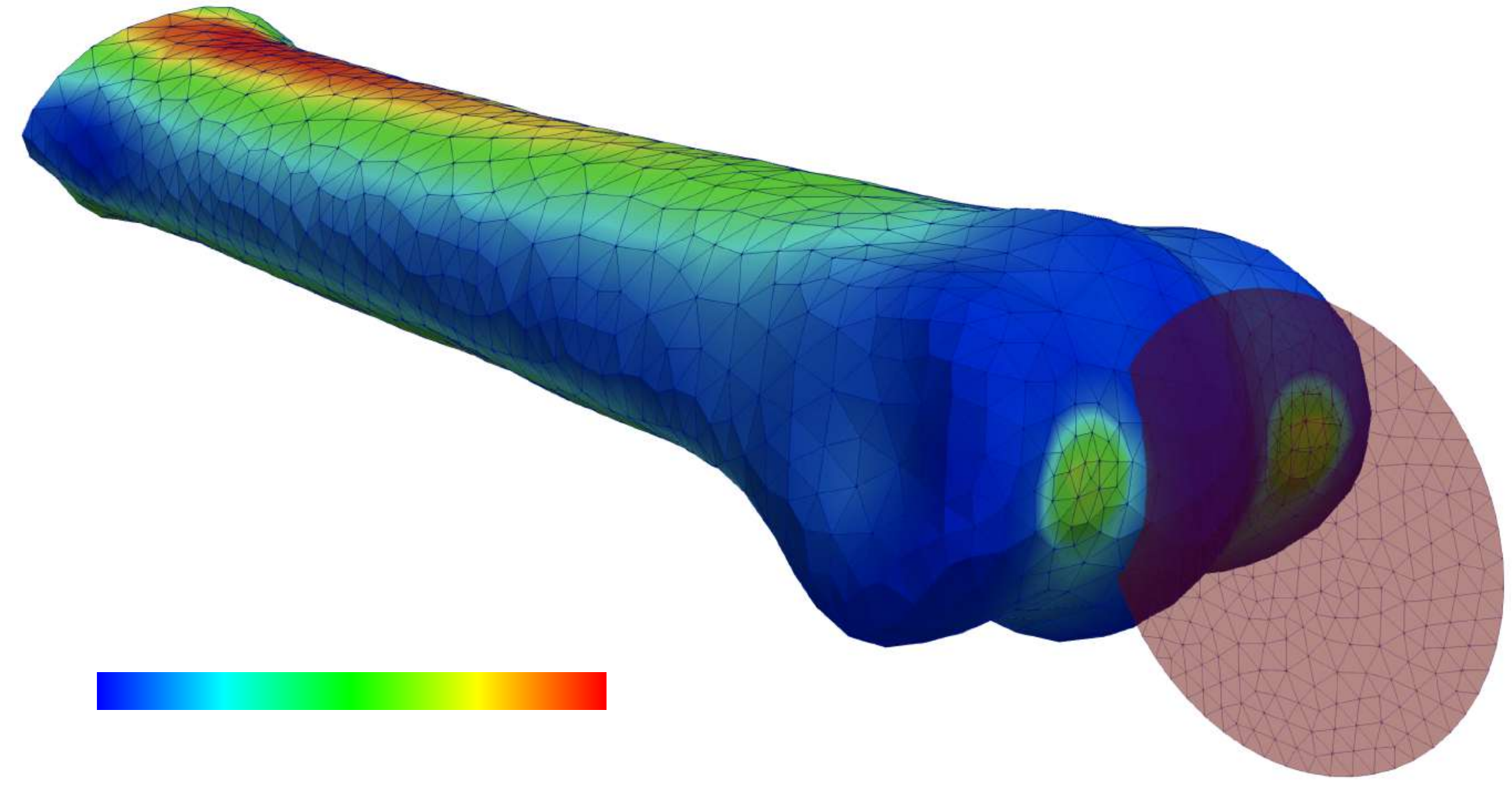
	\caption{Density distribution from bone adaptation analysis mapped onto fracture analysis mesh. Initial crack created using the cutting plane shown.}
	\label{fig:frackmeshcutting}
\end{figure}

\begin{figure}[h!]
	\centering
		\begin{tikzpicture}
			\node at(0,0){	
%
%
\definecolor{mycolor1}{rgb}{1.00000,0.00000,1.00000}%
\begin{tikzpicture}

\begin{axis}[%
width=12cm,
height=8cm,
at={(1.011in,0.723in)},
scale only axis,
xmin=0,
xmax=401,
xlabel style={font=\color{white!15!black}},
xlabel={Time [d]},
ymin=0,
ymax=10,
ylabel style={font=\color{white!15!black}},
ylabel={Fracture energy release rate [$\mathrm{kN} / \mathrm{m}$]},
axis background/.style={fill=white},
legend style={at={(0.97,0.03)}, anchor=south east, legend cell align=left, align=left, fill=none, draw=none}
]
\addplot [color=blue, line width=2.0pt, mark size=2.0pt, mark=triangle, mark options={solid, blue}]
  table[row sep=crcr]{%
0	1.3605\\
5	1.5193\\
25	2.3897\\
40	3.7349\\
75	7.1412\\
97	7.734\\
200	8.2094\\
300	8.2419\\
400	8.2467\\
};
\addlegendentry{$p_{\rm l}= 1$}

\addplot [color=green, line width=2.0pt, mark size=2.0pt, mark=o, mark options={solid, green}]
  table[row sep=crcr]{%
0	1.3764\\
5	1.5382\\
25	2.4522\\
40	3.8552\\
75	7.4199\\
97	8.0468\\
200	8.5459\\
300	8.5798\\
400	8.5847\\
};
\addlegendentry{$p_{\rm l}= 2$}

\addplot [color=mycolor1, line width=2.0pt, mark size=2.0pt, mark=+, mark options={solid, mycolor1}]
  table[row sep=crcr]{%
0	1.38\\
5	1.5426\\
25	2.4674\\
40	3.8853\\
75	7.5005\\
97	8.1385\\
200	8.6484\\
300	8.6831\\
400	8.6882\\
};
\addlegendentry{$p_{\rm l}= 4$}

\end{axis}

\end{tikzpicture}%
};
			\node at(0,0){ \includegraphics[width=9cm]{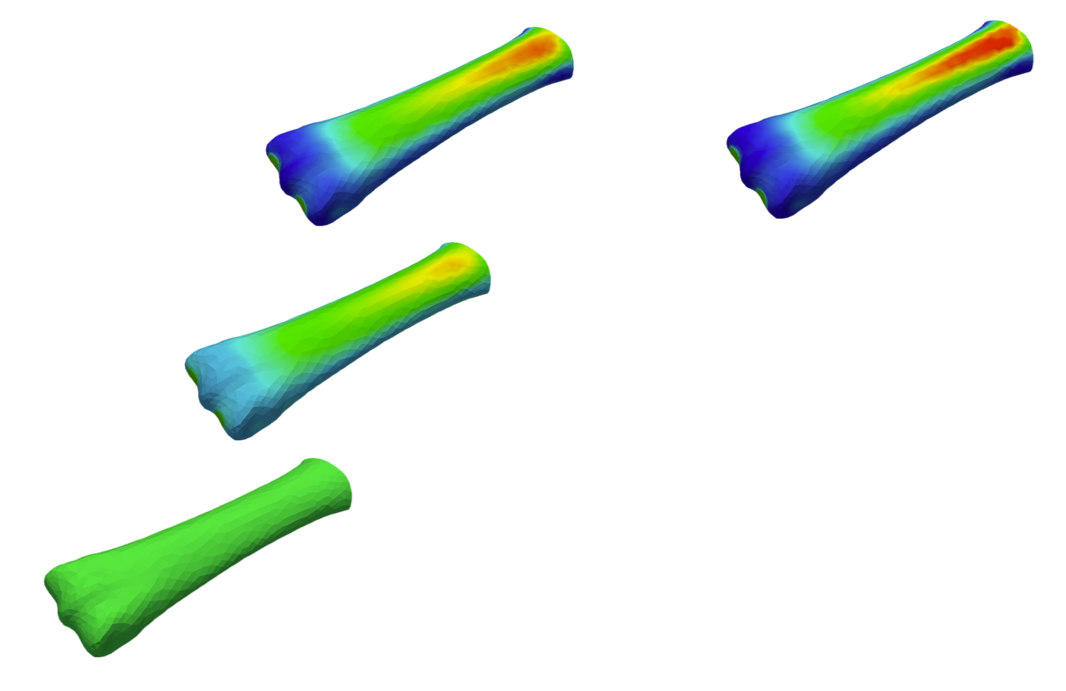}};
			\node at(2,-1){                    	\def\svgwidth{4cm} 	\input{Figures/scale_bars/scale_bar_density_rainbow.pdf_tex}  } ;
		
		\end{tikzpicture}
	\caption{Fracture energy release rate over time during bone adaptation for three local $p$~-~refinements.}
	\label{fig:crackmc3release}
\end{figure}

The previous example is extended to investigate the likelihood of fracture in an equine metacarpal bone at different phases of adaptation during training.
However, this time, densities from a bone adaptation analysis (Section~\ref{sec:numerical_examples:bone_adap}) are mapped onto the coarse mesh, as shown in Figure~\ref{fig:frackmeshcutting}. 
The resulting energy release rate at different points in time of bone adaptation are illustrated in Figure~\ref{fig:crackmc3release} for three different local $p$~-~refinements. It can be seen that the variation in energy release rate for increased orders of approximation at the crack front is very small. 
It can be seen that there is a trend of increasing release energy rate over time and that by introducing a notch in the resorption zone, where no loading is applied, the configurational force attains larger values. This indicates that over time the bone becomes more prone to fracture in this specific region. 

\subsection{Crack propagation in bone}
\begin{figure}[h!]
	\centering
	\begin{tikzpicture}
	\node at(0.,0){  	\input{Figures/tikz/fracture/remodel_frac_compar.tex} };
	\node at (0.25,1.8) {	\includegraphics[width=7cm]{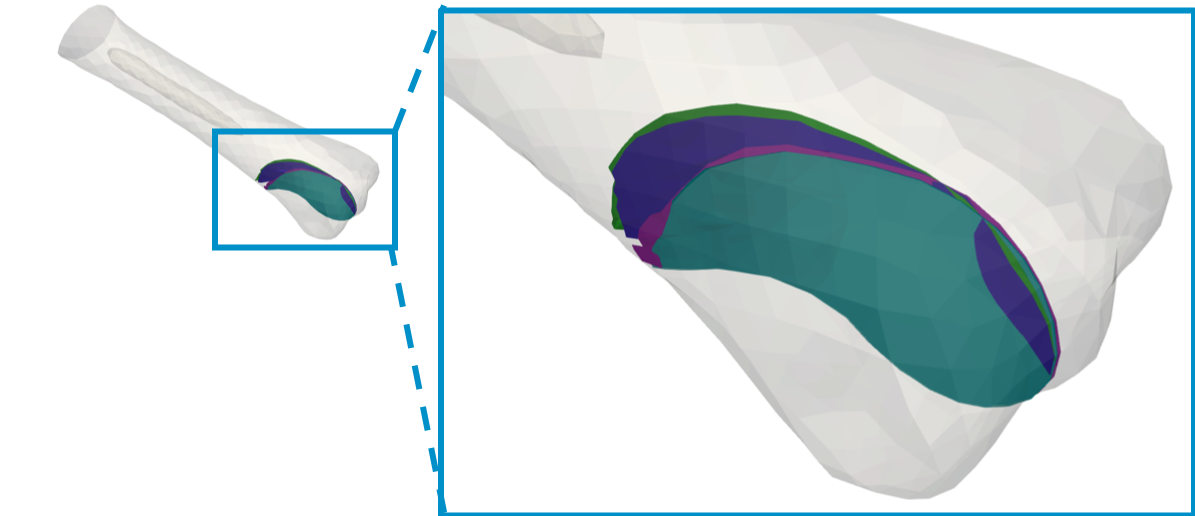}};
	\node at (1.5,0) {	Crack surfaces};
	\end{tikzpicture}
	\caption{Load factor versus crack area for different moments in time during bone adaptation analysis. Bone density distribution influenced both load factor and the resulting crack surface.}
	\label{fig:crack_remodel_frac_compar}
\end{figure}

\begin{figure}[h!]
	\begin{centering}
			\begin{tikzpicture}
				\node at (0,0) {			\includegraphics[width=12cm]{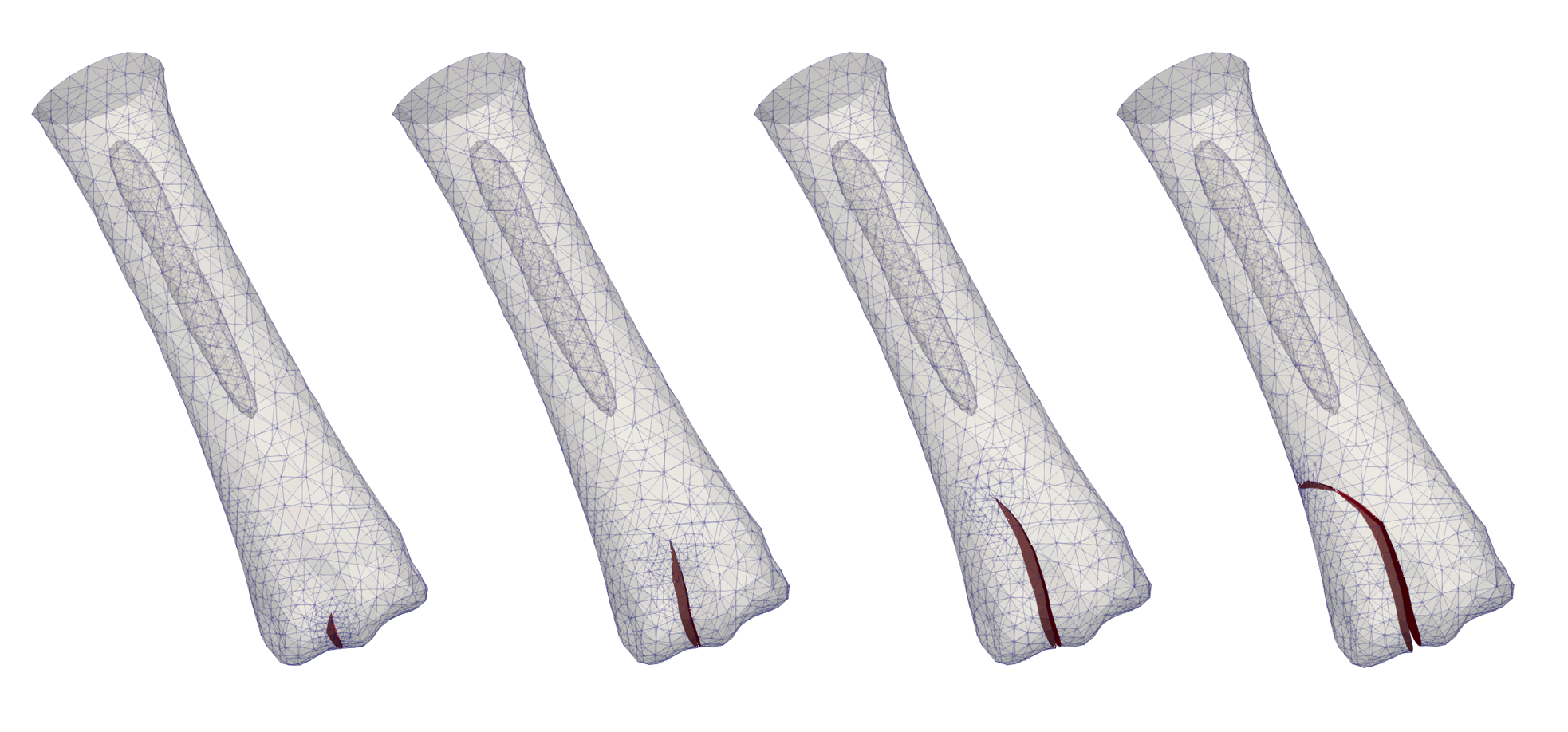} };
				\node at(1,-2.5){a) \hspace{2.5cm} b)  \hspace{2.5cm}  c)  \hspace{2.5cm}  d)  \hspace{2.5cm}  };
		\end{tikzpicture}
		
	\end{centering}		
		\caption{Crack surface evolution in equine 3rd metacarpal.}
		\label{fig:crack_snapshots}
\end{figure}

In this section we move the analysis further by simulating the process of crack propagation for different levels of bone adaptation. The magnitude of applied forces (Figure \ref{fig:mc3_BC}) is controlled by the increment in crack area during each load step using an arc-length technique. The initial finite element mesh is the same as previously, although it is locally refined as the crack front advances. The fracture energy is $2.0\,[\mathrm{ kJ/m^2}]$ for the entire domain. 
All five cases (time snapshots) are solved using 2nd-order approximation functions.
The numerically predicted crack paths are shown in Figure~\ref{fig:crack_snapshots}. 
It can be seen that the crack has an initially planar shape and then curves towards the lateral side of the bone. This simulated crack path compares well with fractures observed in radiographs \citep{whitton2010third}, especially considering the simplified loading conditions. The load factor versus crack area plots are shown in Figure~\ref{fig:crack_remodel_frac_compar}. Consistent with the previous analysis in Section~\ref{sec:mc3_release_eng}, the metacarpal bone shows a decreased resistance to fracture - i.e. for the same crack area, the remodelled bone requires much lower force (load factor) to induce crack propagation. Low density levels at biological equilibrium ($t=90$ and $t=200$) also influences the crack path, with the crack curving earlier than in the initial stages of remodelling. 

\begin{figure}[h]
	\centering
	\begin{minipage}{.45\textwidth}
		\hspace{-1cm}
	\begin{tikzpicture}
		\node at(0.,0){  	
%
%
\begin{tikzpicture}

\begin{axis}[%
width=6cm,
height=6cm,
at={(1.011in,0.723in)},
scale only axis,
xmin=0,
xmax=20,
xlabel style={font=\color{white!15!black}},
xlabel={Crack area $ \rm [cm^2]$},
ymin=0,
ymax=6,
ylabel style={font=\color{white!15!black}},
ylabel={Load factor $\lambda$ [-]},
axis background/.style={fill=white},
legend style={legend cell align=left, align=left, fill=none, draw=none}
]
\addplot [color=blue, line width=2.0pt, mark size=2.0pt, mark=triangle, mark options={solid, blue}]
  table[row sep=crcr]{%
0.11331	4.996\\
0.11511	5.18\\
0.11638	5.454\\
0.1185	5.347\\
0.12091	5.253\\
0.12559	5.282\\
0.12657	5.316\\
0.12699	5.089\\
0.12803	5.076\\
0.12906	5.029\\
0.13096	5.273\\
0.13202	5.214\\
0.13309	5.157\\
0.13255	5.155\\
0.13384	5.139\\
0.13534	5.102\\
0.1362	5.456\\
0.13753	5.361\\
0.14009	5.36\\
0.14251	5.26\\
0.14564	5.168\\
0.15064	5.022\\
0.15371	4.922\\
0.16287	4.561\\
0.17427	4.446\\
0.18363	4.38\\
0.18835	4.647\\
0.19094	4.644\\
0.19762	4.717\\
0.20122	4.754\\
0.20524	4.783\\
0.20837	4.84\\
0.21396	4.974\\
0.22072	5.075\\
0.22226	5.114\\
0.23	5.104\\
0.23928	5.104\\
0.25802	5.001\\
0.27282	5.057\\
0.28937	5.119\\
0.31908	5.158\\
0.34641	5.208\\
0.38505	5.282\\
0.47397	5.328\\
0.52457	5.373\\
0.58961	5.437\\
0.7379	5.499\\
0.83464	5.584\\
0.93194	5.617\\
1.112	5.526\\
1.2139	5.022\\
1.2154	5.084\\
1.2157	5.189\\
1.2179	5.246\\
1.2205	5.274\\
1.2234	5.162\\
1.2276	5.278\\
1.2336	5.318\\
1.2465	5.352\\
1.2608	5.383\\
1.2834	5.401\\
1.3392	5.49\\
1.4041	5.498\\
1.5025	5.486\\
1.6966	5.267\\
1.6986	5.025\\
1.7031	5.057\\
1.7067	4.913\\
1.7092	4.953\\
1.7127	4.975\\
1.7137	4.903\\
1.7199	4.998\\
1.7288	5.041\\
1.7332	5.084\\
1.7529	5.087\\
1.7888	5.085\\
1.8826	5.115\\
1.9555	5.07\\
2.0545	5.046\\
2.2506	4.949\\
2.3501	4.885\\
2.4499	4.829\\
2.6382	4.727\\
2.7374	4.683\\
2.8366	4.632\\
2.9563	4.61\\
2.9862	4.608\\
3.0591	4.654\\
3.1334	4.668\\
3.2327	4.65\\
3.4255	4.353\\
3.5247	4.314\\
3.6241	4.274\\
3.8231	4.161\\
3.9228	4.108\\
4.0226	4.059\\
4.2016	4.036\\
4.3015	4.001\\
4.4015	3.957\\
4.5981	3.852\\
4.6983	3.812\\
4.7986	3.77\\
5.0002	3.671\\
5.1006	3.644\\
5.2011	3.607\\
5.3998	3.513\\
5.4996	3.486\\
5.5996	3.458\\
5.7784	3.376\\
5.8784	3.354\\
5.9783	3.318\\
6.1884	3.253\\
6.2886	3.221\\
6.3884	3.192\\
6.4632	3.199\\
6.4989	3.202\\
6.6328	3.228\\
6.7331	3.232\\
6.8333	3.187\\
7.0398	3.052\\
7.1402	3.024\\
7.2405	2.995\\
7.4362	2.895\\
7.5365	2.857\\
7.6366	2.821\\
7.8331	2.779\\
7.9335	2.747\\
8.0339	2.72\\
8.1477	2.776\\
8.1873	2.792\\
8.2924	2.842\\
8.3924	2.82\\
8.4925	2.747\\
8.6787	2.665\\
8.7791	2.618\\
8.8796	2.585\\
9.0797	2.489\\
9.1798	2.454\\
9.2799	2.422\\
9.3813	2.46\\
9.3942	2.459\\
9.4099	2.473\\
9.4322	2.47\\
9.476	2.5\\
9.5255	2.517\\
9.6039	2.529\\
9.776	2.44\\
9.8768	2.359\\
9.9778	2.3\\
10.175	2.205\\
10.277	2.18\\
10.377	2.158\\
10.565	2.114\\
10.66	2.062\\
10.742	2.046\\
10.931	2.022\\
11.021	2.004\\
11.12	1.992\\
11.3	1.893\\
11.391	1.885\\
11.492	1.85\\
11.671	1.832\\
11.772	1.824\\
11.872	1.812\\
12.073	1.717\\
12.174	1.708\\
12.274	1.697\\
12.468	1.674\\
12.539	1.683\\
12.638	1.646\\
12.835	1.596\\
12.935	1.584\\
13.035	1.572\\
13.223	1.513\\
13.323	1.497\\
13.423	1.483\\
13.608	1.491\\
13.688	1.493\\
13.788	1.462\\
13.969	1.385\\
14.047	1.387\\
14.148	1.359\\
14.272	1.335\\
14.296	1.322\\
14.346	1.319\\
14.398	1.331\\
14.493	1.349\\
14.689	1.315\\
14.789	1.283\\
14.89	1.256\\
15.085	1.229\\
15.185	1.213\\
15.286	1.192\\
15.478	1.123\\
15.579	1.102\\
15.68	1.081\\
15.857	1.008\\
15.939	1.012\\
16.04	0.9733\\
16.228	0.9232\\
16.311	0.921\\
16.402	0.8839\\
16.574	0.8075\\
16.629	0.8217\\
16.729	0.7629\\
16.93	0.6952\\
16.98	0.6857\\
17.058	0.6572\\
17.189	0.5898\\
17.279	0.5587\\
17.347	0.5531\\
17.46	0.458\\
17.563	0.423\\
17.654	0.3933\\
17.799	0.3173\\
17.903	0.2813\\
17.981	0.3075\\
18.071	0.1716\\
18.113	0.1657\\
18.141	0.179\\
18.161	0.1411\\
18.15	0.1111\\
18.157	0.1113\\
18.119	0.1035\\
18.132	0.1005\\
18.15	0.09471\\
18.176	0.07829\\
18.218	0.06477\\
18.277	0.04669\\
};
\addlegendentry{normal}

\addplot [color=green, line width=2.0pt, mark size=2.0pt, mark=o, mark options={solid, green}]
  table[row sep=crcr]{%
0.1072	5.215\\
0.10938	5.038\\
0.10066	5.313\\
0.10417	5.208\\
0.10147	5.535\\
0.10267	5.508\\
0.10427	5.474\\
0.10541	5.4\\
0.10785	5.395\\
0.11133	5.409\\
0.11855	5.174\\
0.12315	5.094\\
0.12864	5.03\\
0.13918	4.695\\
0.14777	4.6\\
0.15532	4.441\\
0.17552	4.553\\
0.19395	4.631\\
0.21609	4.77\\
0.26323	5.158\\
0.26902	5.331\\
0.27942	5.037\\
0.29223	5.126\\
0.3102	5.237\\
0.34189	5.297\\
0.37212	5.359\\
0.4081	5.457\\
0.48271	5.518\\
0.52849	5.54\\
0.57447	5.59\\
0.66654	5.564\\
0.68031	5.691\\
0.71242	5.744\\
0.75747	5.958\\
0.82808	6.136\\
0.99086	5.637\\
1.0879	5.543\\
1.1856	5.523\\
1.3788	5.307\\
1.4768	5.247\\
1.4958	5.402\\
1.5411	5.516\\
1.6236	5.654\\
1.8025	5.437\\
1.9007	5.311\\
1.9906	5.202\\
2.1744	5.093\\
2.2735	5.037\\
2.3646	5.003\\
2.4602	4.882\\
2.4955	4.874\\
2.5894	4.863\\
2.6801	4.877\\
2.7794	4.828\\
2.9702	4.674\\
3.0694	4.604\\
3.1686	4.525\\
3.3671	4.281\\
3.4668	4.238\\
3.5666	4.208\\
3.7579	4.104\\
3.8577	4.065\\
3.9576	4.037\\
4.1477	4.027\\
4.2478	4.005\\
4.348	3.977\\
4.5337	3.871\\
4.6334	3.839\\
4.733	3.816\\
4.9396	3.606\\
5.0403	3.585\\
5.1406	3.56\\
5.3386	3.477\\
5.3636	3.552\\
5.4287	3.643\\
5.5008	3.67\\
5.6008	3.666\\
5.7822	3.468\\
5.8825	3.421\\
5.9827	3.371\\
6.1659	3.285\\
6.2664	3.257\\
6.3668	3.246\\
6.5775	3.204\\
6.6783	3.175\\
6.7785	3.143\\
6.943	3.055\\
7.0468	3.051\\
7.1495	3.034\\
7.2666	2.991\\
7.3122	3.026\\
7.4335	3.047\\
7.5345	3.023\\
7.6352	2.972\\
7.8259	2.877\\
7.9261	2.854\\
8.0262	2.832\\
8.1974	2.689\\
8.2974	2.665\\
8.3977	2.646\\
8.6078	2.531\\
8.7088	2.511\\
8.81	2.493\\
8.9312	2.484\\
8.9596	2.463\\
9.0285	2.483\\
9.0921	2.483\\
9.1917	2.497\\
9.3848	2.364\\
9.4863	2.31\\
9.5875	2.277\\
9.785	2.266\\
9.8846	2.246\\
9.9847	2.227\\
10.171	2.117\\
10.269	2.085\\
10.376	1.977\\
10.38	2.014\\
10.378	2.02\\
10.383	2.022\\
10.388	2.022\\
10.403	2.021\\
10.41	2.026\\
10.417	2.03\\
10.427	2.033\\
10.438	2.037\\
10.453	2.04\\
10.494	2.037\\
10.52	2.042\\
10.557	2.053\\
10.641	2.056\\
10.708	2.056\\
10.779	2.054\\
10.687	1.98\\
10.69	1.987\\
10.695	1.992\\
10.701	2.003\\
10.71	2.016\\
10.724	1.79\\
10.748	1.803\\
10.791	1.819\\
10.895	2.048\\
10.991	2.059\\
11.091	2.038\\
11.276	1.958\\
11.377	1.955\\
11.398	1.965\\
11.444	2.009\\
11.516	2.028\\
11.671	2.023\\
11.771	1.948\\
11.872	1.911\\
12.01	1.842\\
12.05	1.866\\
12.156	1.925\\
12.237	1.886\\
12.337	1.852\\
12.528	1.7\\
12.553	1.762\\
12.6	1.869\\
12.665	1.857\\
12.765	1.872\\
12.945	1.667\\
13.045	1.631\\
13.145	1.602\\
13.339	1.562\\
13.43	1.558\\
13.531	1.504\\
13.648	1.52\\
13.688	1.532\\
13.775	1.489\\
13.856	1.477\\
13.956	1.466\\
14.072	1.575\\
14.118	1.609\\
14.237	1.549\\
14.337	1.516\\
14.363	1.686\\
14.402	1.658\\
14.465	1.637\\
14.609	1.565\\
14.71	1.466\\
14.81	1.4\\
15.004	1.288\\
15.105	1.26\\
15.206	1.239\\
15.304	1.238\\
15.339	1.195\\
15.384	1.238\\
15.409	1.264\\
15.505	1.271\\
15.57	1.233\\
15.67	1.214\\
15.859	1.134\\
15.959	1.088\\
16.06	1.054\\
16.186	1.026\\
16.216	1.001\\
16.272	0.9948\\
16.287	1.042\\
16.321	1.09\\
16.364	1.084\\
16.427	1.09\\
16.568	0.9815\\
16.669	0.9306\\
16.77	0.9005\\
16.873	0.8732\\
16.912	0.8676\\
17.011	0.8442\\
17.096	0.8255\\
17.088	0.865\\
17.129	0.8665\\
17.192	0.8666\\
17.336	0.7717\\
17.437	0.7204\\
17.539	0.6751\\
17.655	0.6473\\
17.691	0.6336\\
17.749	0.6084\\
17.81	0.5756\\
17.882	0.5537\\
17.99	0.4469\\
18.008	0.4545\\
18.034	0.4277\\
18.072	0.4087\\
18.126	0.389\\
18.201	0.3361\\
18.27	0.3021\\
18.269	0.3042\\
18.293	0.296\\
18.329	0.2863\\
18.383	0.2506\\
18.439	0.2254\\
18.505	0.2027\\
18.576	0.1608\\
18.591	0.1677\\
18.584	0.135\\
18.609	0.1237\\
18.644	0.1142\\
18.61	0.07903\\
18.613	0.07949\\
18.557	0.07669\\
18.565	0.07496\\
18.576	0.07236\\
18.608	0.06696\\
18.626	0.06211\\
18.649	0.05618\\
18.666	0.04152\\
18.691	0.03502\\
18.715	0.02921\\
};
\addlegendentry{with singularity}

\end{axis}
\end{tikzpicture}%
 };
	\end{tikzpicture}
\end{minipage}%
\quad
\begin{minipage}{.45\textwidth}
	\begin{tikzpicture}
	\node at(0.,0){  	
%
%
\begin{tikzpicture}

\begin{axis}[%
width=6cm,
height=6cm,
at={(1.011in,0.723in)},
restrict y to domain=0:20,
scale only axis,
xmin=0,
xmax=20,
xlabel style={font=\color{white!15!black}},
xlabel={Crack area $ \rm [cm^2]$},
ymin=0,
ymax=6,
ylabel style={font=\color{white!15!black}},
ylabel={Load factor $\lambda$ [-]},
axis background/.style={fill=white},
legend style={legend cell align=left, align=left, fill=none, draw=none}
]
\addplot [color=green, line width=2.0pt, mark size=2.0pt, mark=triangle, mark options={solid, green}]
  table[row sep=crcr]{%
0.11338	2.635\\
0.11551	2.777\\
0.11675	2.923\\
0.11757	2.914\\
0.11865	2.893\\
0.11951	2.882\\
0.12043	2.867\\
0.11993	2.883\\
0.12089	2.865\\
0.12186	2.84\\
0.12266	3.038\\
0.12365	3.001\\
0.12514	3.166\\
0.12666	2.976\\
0.12837	2.947\\
0.12944	2.808\\
0.1319	2.676\\
0.13407	2.644\\
0.13744	2.672\\
0.13847	2.861\\
0.13984	2.734\\
0.14177	2.712\\
0.14424	2.69\\
0.14173	2.729\\
0.14422	2.673\\
0.1424	2.803\\
0.14384	2.789\\
0.14587	2.774\\
0.14609	2.66\\
0.1498	2.635\\
0.15419	2.611\\
0.1527	2.905\\
0.15457	2.916\\
0.1535	2.843\\
0.15783	2.807\\
0.16461	2.82\\
0.17411	2.738\\
0.18784	2.725\\
0.20544	2.676\\
0.24416	2.659\\
0.27829	2.709\\
0.32667	2.767\\
0.44303	2.801\\
0.52505	2.865\\
0.62054	2.938\\
0.75284	2.906\\
0.78035	2.957\\
0.84701	3.165\\
0.9024	3.132\\
0.98956	3.135\\
1.1721	3.171\\
1.2702	3.142\\
1.3685	3.122\\
1.5601	3.039\\
1.6583	3.008\\
1.7565	2.985\\
1.9376	2.925\\
2.0364	2.909\\
2.1353	2.891\\
2.3292	2.766\\
2.4283	2.742\\
2.5275	2.715\\
2.6099	2.483\\
2.621	2.471\\
2.6221	2.485\\
2.6231	2.49\\
2.623	2.494\\
2.6241	2.498\\
2.6255	2.504\\
2.6162	2.451\\
2.6177	2.503\\
2.6193	2.512\\
2.618	2.53\\
2.6191	2.532\\
2.6202	2.535\\
2.6152	2.542\\
2.6163	2.546\\
2.6175	2.55\\
2.6212	2.487\\
2.6224	2.495\\
2.6234	2.504\\
2.6214	2.501\\
2.6224	2.507\\
2.6234	2.516\\
2.6128	2.519\\
2.6138	2.527\\
2.6148	2.529\\
2.6259	2.573\\
2.6269	2.58\\
2.6281	2.58\\
2.6333	2.568\\
2.6345	2.576\\
2.6359	2.579\\
2.6393	2.574\\
2.6411	2.576\\
2.6434	2.578\\
2.6468	2.582\\
2.6503	2.587\\
2.6548	2.592\\
2.6094	2.623\\
2.6185	2.634\\
2.6303	2.645\\
2.6792	2.639\\
2.7055	2.66\\
2.7534	2.675\\
2.8858	2.728\\
2.9844	2.706\\
3.083	2.67\\
3.284	2.616\\
3.3835	2.582\\
3.4829	2.554\\
3.6783	2.419\\
3.7779	2.399\\
3.8774	2.378\\
4.074	2.408\\
4.1739	2.389\\
4.2739	2.366\\
4.451	2.296\\
4.5512	2.289\\
4.6513	2.266\\
4.854	2.215\\
4.954	2.193\\
5.0539	2.172\\
5.2512	2.08\\
5.3512	2.063\\
5.4512	2.048\\
5.6495	1.999\\
5.7496	1.981\\
5.8496	1.965\\
6.0468	1.933\\
6.1468	1.916\\
6.2468	1.897\\
6.4366	1.876\\
6.5364	1.857\\
6.6365	1.839\\
6.829	1.799\\
6.9294	1.782\\
7.0299	1.767\\
7.2155	1.73\\
7.3154	1.714\\
7.4153	1.698\\
7.6104	1.659\\
7.7104	1.642\\
7.8104	1.625\\
8.0109	1.574\\
8.1063	1.561\\
8.2064	1.547\\
8.4095	1.501\\
8.5008	1.485\\
8.601	1.469\\
8.7966	1.431\\
8.8969	1.42\\
8.9972	1.41\\
9.1825	1.345\\
9.2824	1.331\\
9.3827	1.318\\
9.4918	1.321\\
9.5198	1.324\\
9.5843	1.327\\
9.6543	1.327\\
9.7549	1.321\\
9.9451	1.307\\
10.046	1.286\\
10.146	1.267\\
10.339	1.242\\
10.441	1.218\\
10.542	1.201\\
10.723	1.148\\
10.824	1.134\\
10.925	1.12\\
11.088	1.001\\
11.188	0.9865\\
11.288	0.973\\
11.493	1.02\\
11.593	1.006\\
11.694	0.9918\\
11.872	0.9786\\
11.974	0.9646\\
12.075	0.9531\\
12.193	0.9376\\
12.233	0.9399\\
12.332	0.9389\\
12.432	0.9362\\
12.533	0.9222\\
12.721	0.881\\
12.821	0.8659\\
12.922	0.8528\\
13.092	0.7999\\
13.193	0.787\\
13.295	0.7768\\
13.468	0.7521\\
13.568	0.7369\\
13.669	0.7225\\
13.851	0.6834\\
13.95	0.6651\\
14.05	0.6481\\
14.236	0.6146\\
14.336	0.5992\\
14.436	0.5843\\
14.621	0.5428\\
14.722	0.5249\\
14.823	0.5073\\
15.003	0.4604\\
15.104	0.4428\\
15.206	0.4269\\
15.384	0.3839\\
15.485	0.3662\\
15.585	0.3485\\
15.785	0.2946\\
15.887	0.2744\\
15.989	0.2547\\
16.116	0.2106\\
16.22	0.1898\\
16.323	0.169\\
16.376	0.1399\\
16.395	0.1319\\
16.408	0.1214\\
16.449	0.114\\
16.514	0.1024\\
16.624	0.07002\\
16.73	0.05029\\
16.808	0.03369\\
};
\addlegendentry{homogeneous}

\addplot [color=red, line width=2.0pt, mark size=2.0pt, mark=o, mark options={solid, red}]
  table[row sep=crcr]{%
0.11331	4.996\\
0.11511	5.18\\
0.11638	5.454\\
0.1185	5.347\\
0.12091	5.253\\
0.12559	5.282\\
0.12657	5.316\\
0.12699	5.089\\
0.12803	5.076\\
0.12906	5.029\\
0.13096	5.273\\
0.13202	5.214\\
0.13309	5.157\\
0.13255	5.155\\
0.13384	5.139\\
0.13534	5.102\\
0.1362	5.456\\
0.13753	5.361\\
0.14009	5.36\\
0.14251	5.26\\
0.14564	5.168\\
0.15064	5.022\\
0.15371	4.922\\
0.16287	4.561\\
0.17427	4.446\\
0.18363	4.38\\
0.18835	4.647\\
0.19094	4.644\\
0.19762	4.717\\
0.20122	4.754\\
0.20524	4.783\\
0.20837	4.84\\
0.21396	4.974\\
0.22072	5.075\\
0.22226	5.114\\
0.23	5.104\\
0.23928	5.104\\
0.25802	5.001\\
0.27282	5.057\\
0.28937	5.119\\
0.31908	5.158\\
0.34641	5.208\\
0.38505	5.282\\
0.47397	5.328\\
0.52457	5.373\\
0.58961	5.437\\
0.7379	5.499\\
0.83464	5.584\\
0.93194	5.617\\
1.112	5.526\\
1.2139	5.022\\
1.2154	5.084\\
1.2157	5.189\\
1.2179	5.246\\
1.2205	5.274\\
1.2234	5.162\\
1.2276	5.278\\
1.2336	5.318\\
1.2465	5.352\\
1.2608	5.383\\
1.2834	5.401\\
1.3392	5.49\\
1.4041	5.498\\
1.5025	5.486\\
1.6966	5.267\\
1.6986	5.025\\
1.7031	5.057\\
1.7067	4.913\\
1.7092	4.953\\
1.7127	4.975\\
1.7137	4.903\\
1.7199	4.998\\
1.7288	5.041\\
1.7332	5.084\\
1.7529	5.087\\
1.7888	5.085\\
1.8826	5.115\\
1.9555	5.07\\
2.0545	5.046\\
2.2506	4.949\\
2.3501	4.885\\
2.4499	4.829\\
2.6382	4.727\\
2.7374	4.683\\
2.8366	4.632\\
2.9563	4.61\\
2.9862	4.608\\
3.0591	4.654\\
3.1334	4.668\\
3.2327	4.65\\
3.4255	4.353\\
3.5247	4.314\\
3.6241	4.274\\
3.8231	4.161\\
3.9228	4.108\\
4.0226	4.059\\
4.2016	4.036\\
4.3015	4.001\\
4.4015	3.957\\
4.5981	3.852\\
4.6983	3.812\\
4.7986	3.77\\
5.0002	3.671\\
5.1006	3.644\\
5.2011	3.607\\
5.3998	3.513\\
5.4996	3.486\\
5.5996	3.458\\
5.7784	3.376\\
5.8784	3.354\\
5.9783	3.318\\
6.1884	3.253\\
6.2886	3.221\\
6.3884	3.192\\
6.4632	3.199\\
6.4989	3.202\\
6.6328	3.228\\
6.7331	3.232\\
6.8333	3.187\\
7.0398	3.052\\
7.1402	3.024\\
7.2405	2.995\\
7.4362	2.895\\
7.5365	2.857\\
7.6366	2.821\\
7.8331	2.779\\
7.9335	2.747\\
8.0339	2.72\\
8.1477	2.776\\
8.1873	2.792\\
8.2924	2.842\\
8.3924	2.82\\
8.4925	2.747\\
8.6787	2.665\\
8.7791	2.618\\
8.8796	2.585\\
9.0797	2.489\\
9.1798	2.454\\
9.2799	2.422\\
9.3813	2.46\\
9.3942	2.459\\
9.4099	2.473\\
9.4322	2.47\\
9.476	2.5\\
9.5255	2.517\\
9.6039	2.529\\
9.776	2.44\\
9.8768	2.359\\
9.9778	2.3\\
10.175	2.205\\
10.277	2.18\\
10.377	2.158\\
10.565	2.114\\
10.66	2.062\\
10.742	2.046\\
10.931	2.022\\
11.021	2.004\\
11.12	1.992\\
11.3	1.893\\
11.391	1.885\\
11.492	1.85\\
11.671	1.832\\
11.772	1.824\\
11.872	1.812\\
12.073	1.717\\
12.174	1.708\\
12.274	1.697\\
12.468	1.674\\
12.539	1.683\\
12.638	1.646\\
12.835	1.596\\
12.935	1.584\\
13.035	1.572\\
13.223	1.513\\
13.323	1.497\\
13.423	1.483\\
13.608	1.491\\
13.688	1.493\\
13.788	1.462\\
13.969	1.385\\
14.047	1.387\\
14.148	1.359\\
14.272	1.335\\
14.296	1.322\\
14.346	1.319\\
14.398	1.331\\
14.493	1.349\\
14.689	1.315\\
14.789	1.283\\
14.89	1.256\\
15.085	1.229\\
15.185	1.213\\
15.286	1.192\\
15.478	1.123\\
15.579	1.102\\
15.68	1.081\\
15.857	1.008\\
15.939	1.012\\
16.04	0.9733\\
16.228	0.9232\\
16.311	0.921\\
16.402	0.8839\\
16.574	0.8075\\
16.629	0.8217\\
16.729	0.7629\\
16.93	0.6952\\
16.98	0.6857\\
17.058	0.6572\\
17.189	0.5898\\
17.279	0.5587\\
17.347	0.5531\\
17.46	0.458\\
17.563	0.423\\
17.654	0.3933\\
17.799	0.3173\\
17.903	0.2813\\
17.981	0.3075\\
18.071	0.1716\\
18.113	0.1657\\
18.141	0.179\\
18.161	0.1411\\
18.15	0.1111\\
18.157	0.1113\\
18.119	0.1035\\
18.132	0.1005\\
18.15	0.09471\\
18.176	0.07829\\
18.218	0.06477\\
18.277	0.04669\\
};
\addlegendentry{heterogeneous}

\end{axis}
\end{tikzpicture}%
 };
	\node at (2,0.35) {	\includegraphics[width=3.5cm]{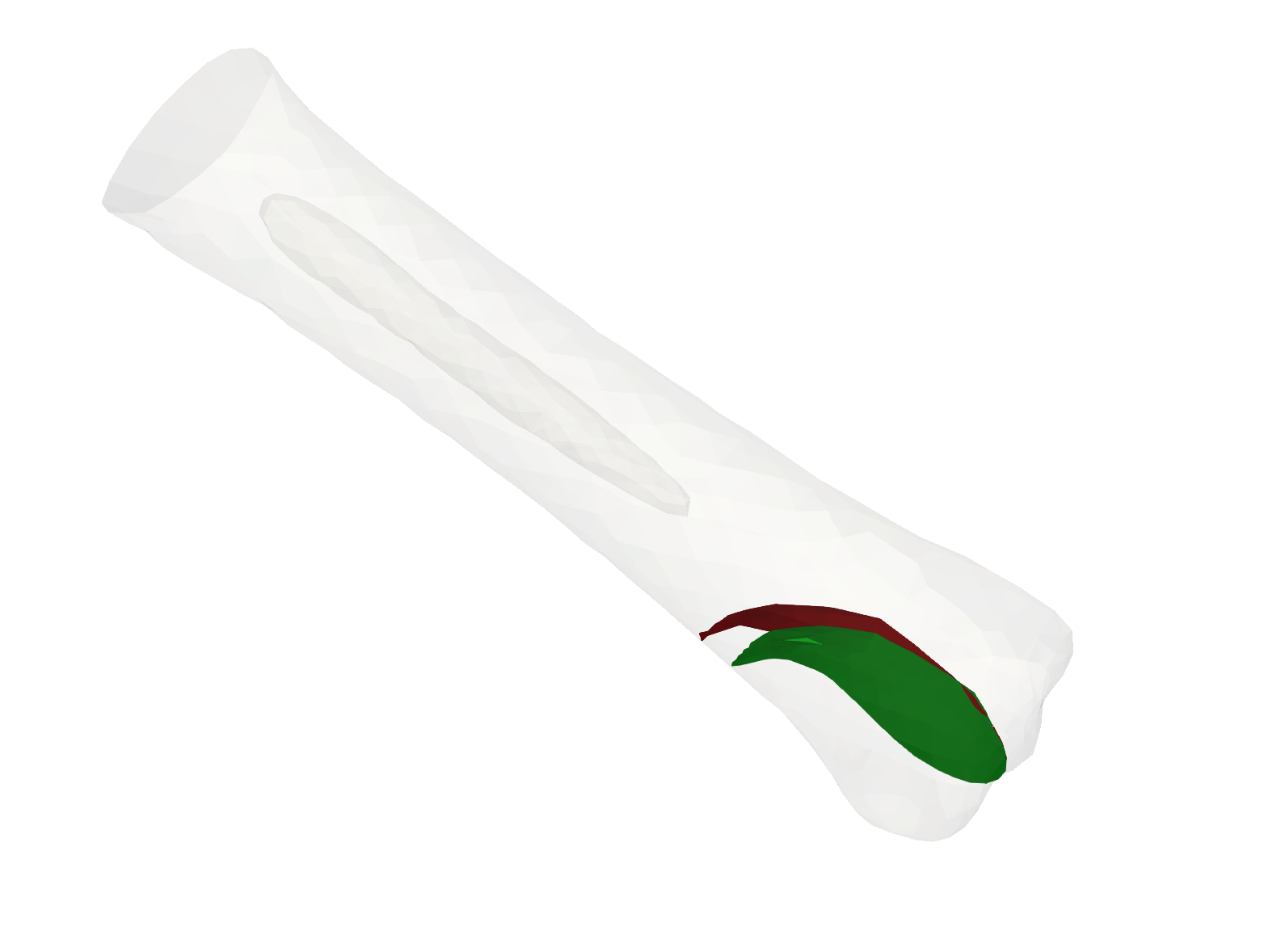}};
	\end{tikzpicture}
\end{minipage}

(a) \hspace{6cm} (b)
	\caption{Load factor versus crack area for (a) with and without singularity element and (b) homogeneous versus heterogeneous density distribution.}
	\label{fig:load_factor1}
\end{figure}

As previously demonstrated (Figure~\ref{fig:plate_conv}(b)) modelling singularity can significantly improve the accuracy of the configurational forces at the crack front. In the next example we considered bone with heterogeneous density distribution mapped from CT scan data. 
In the Figure~\ref{fig:load_factor1}(a) results from crack analysis with and without Quarter Point elements are depicted. It is evident that accurate stress state at the tip has a negligible impact on the full crack propagation analysis and the resulting load factor.  

From the load-crack area curves in Figure~\ref{fig:load_factor1}(b) it can be observed that including density data from CT scans have a significant impact on the predicted load factor and crack path as well.

\begin{figure}[h]
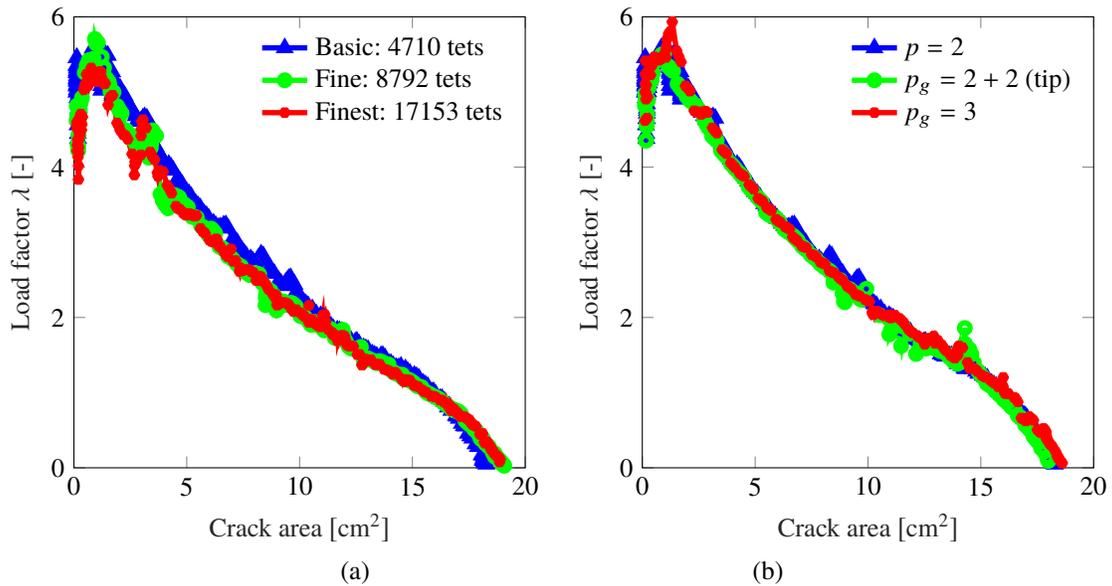

	\centering
	\begin{minipage}{.5\textwidth}
		\begin{tikzpicture}
			\node at(0.,0){ \input{Figures/tikz/fracture/real_fract_ct_h_ref.tex} };
		 \end{tikzpicture}
	\end{minipage}%
	\begin{minipage}{.5\textwidth}
		\begin{tikzpicture}
			\node at(0,0){         \input{Figures/tikz/fracture/real_fract_ct_p_ref.tex} };
		\end{tikzpicture}
	\end{minipage}
	
	(a)\hspace{5cm} (b)
			 \caption{Load factor versus crack area (a) $h$~-~refinement). (b) $p$~-~refinement.}
		 \label{fig:load_factor2}
	\end{figure}

Finally, we investigated the h and p convergence. The results presented on the Figures \ref{fig:load_factor2}(a) and \ref{fig:load_factor2}(b) show good numerical convergence for consecutive refinements. It can be concluded that our formulation predicts crack path accurately with minimal effect from original mesh or order of approximation. 

\section{Discussion}\label{sec:discussion}
This paper has presented a FEM computational modelling framework to investigate the influence of bone adaptation, and associated bone density distribution, on fracture resistance and fracture propagation. 
The influence of the heterogeneous density distribution was captured using an extension of the authors' previous work on configurational mechanics for fracture. 
Configurational forces are the driver for crack propagation and it was shown that in order to evaluate correctly these forces at the crack front it is necessary to have a spatially smooth density field, with higher regularity than if the field is directly approximated on the finite element mesh. 
Therefore, density data is approximated as a smooth field using a Moving Weighted Least Squares method. 
In this paper, the bone density field was generated from both bone adaptation analyses and from subject-specific geometry and material properties obtained from CT scans. 
It is important to note that the adoption of configurational mechanics avoids the need for post-processing, since configurational forces, and the fracture energy release rate, are expressed exclusively in terms of nodal quantities.

The constitutive model for bone adaptation included a bell function to define the rate of adaptation. 
This did not enforce rigid bounds on density levels, but merely slowed down the rate of convergence to biological equilibrium. 
This approach will be useful when trying fit model parameters to the actual density data form CT scans in defined periods of time. 
It is also possible to enforce bounds on density levels by introducing and calibrating mass influx in the mass balance equation~\citep{sharma2013adaptive}. 

Numerical examples demonstrated the performance and accuracy of the proposed framework. 
Numerical convergence was demonstrated for all examples and the use of singularity elements was shown to further improve the rate of convergence. However, it was also confirmed that improved accuracy of the stress at the tip had no impact on the crack propagation analysis and the resulting crack path.  The final example, demonstrated how mechanical loading and subsequent adaptation influence the resistance to bone fracture. Therefore, this framework will be a useful tool in understanding fractures in bone and ultimately preventing catastrophic fractures. 

All analyses were undertaken using MoFEM~\citep{mofem2017} that has been developed to support scalability and ensure robustness. The entire framework can be executed on parallel computer systems. Supplementary data (CT scans, mesh files, command lines) necessary to reproduce the results of all numerical examples can be found in~\citep{karol_lewandowski_2019_dataset}. The bone adaptation and fracture mechanics are both submodules in the MoFEM library~\citep{mofem2017}, which can be installed using the flexible package manager, Spack~\citep{spack2015}.

\newpage
\bibliography{bibfile.bib}

\end{document}